\documentclass[iop,revtex4]{emulateapj}
\usepackage{amsmath}
\pdfoutput=1

\providecommand{\HII}{H~{\textsc i}{\textsc i}}    
\providecommand{\OIII}{[O~{\textsc i}{\textsc i}{\textsc i}]}  
\providecommand{\OII}{[O~{\textsc i}{\textsc i}]}    
\providecommand{\SIII}{[S~{\textsc i}{\textsc i}{\textsc i}]}  
\providecommand{\SII}{[S~{\textsc i}{\textsc i}]}    
\providecommand{\NII}{[N~{\textsc i}{\textsc i}]}    
\providecommand{\NeIII}{[Ne~{\textsc i}{\textsc i}{\textsc i}]} 
\providecommand{\NI}{[N~{\textsc i}]}     
\providecommand{\OI}{[O~{\textsc i}]}     
\providecommand{\ArIII}{[Ar~{\textsc i}{\textsc i}{\textsc i}]}  
\providecommand{\ArIV}{[Ar~{\textsc i}{\textsc v}]}   
\providecommand{\HA}{H$\alpha$}      
\providecommand{\HB}{H$\beta$}      
\providecommand{\HG}{H$\gamma$}     
\providecommand{\HD}{H$\delta$}      
\providecommand{\HE}{H$\epsilon$}      
\providecommand{\R}{$R_{23}$}      
\providecommand{\Te}{$T_e$}       
\providecommand{\cHB}{c$_{\text{H}\beta}$}    
\providecommand{\abun}{12+log(O/H)}     

\shorttitle{Nebular Abundances of Fifteen Intermediate Luminosity Star-Forming Galaxies}
\shortauthors{Hirschauer et al.}

\begin{document}


\title{Metal Abundances of KISS Galaxies. V. Nebular Abundances of Fifteen Intermediate Luminosity Star-Forming Galaxies}







\author{Alec S. Hirschauer$^1$, John J. Salzer$^1$, Fabio Bresolin$^2$, Ivo Saviane$^3$, Irina Yegorova$^3$}
\affil{$^1$Astronomy Department, Indiana University, Bloomington, IN 47405}
\affil{$^2$Institute for Astronomy, University of Hawaii, Honolulu, HI 96822}
\affil{$^3$European Southern Observatory, Alonso de Cordova 3107, Santiago, Chile}
\email{ash@astro.indiana.edu}

\begin{abstract}
We present high S/N spectroscopy of 15 emission-line galaxies (ELGs) cataloged in the KPNO International Spectroscopic Survey (KISS), selected for their possession of high equivalent width \OIII\ lines.
The primary goal of this study was to attempt to derive direct-method (\Te) abundances for use in constraining the upper-metallicity branch of the \R\ relation.
The spectra cover the full optical region from \OII$\lambda\lambda$3726,3729 to \SIII$\lambda\lambda$9069,9531 and include the measurement of \OIII$\lambda$4363 in 13 objects.
From these spectra, we determine abundance ratios of helium, nitrogen, oxygen, neon, sulfur, and argon.
We find these galaxies to predominantly possess oxygen abundances in the range of 8.0 $\lesssim$ \abun\ $\lesssim$ 8.3.
We present a comparison of direct-method abundances with empirical SEL techniques, revealing several discrepancies.
We also present a comparison of direct-method oxygen abundance calculations using electron temperatures determined from emission lines of O$^{++}$ and S$^{++}$, finding a small systematic shift to lower \Te\ ($\sim$1184 K) and higher metallicity ($\sim$0.14 dex) for sulfur-derived \Te\ compared to oxygen-derived \Te.
Finally, we explore in some detail the different spectral activity types of targets in our sample, including regular star-forming galaxies, those with suspected AGN contamination, and a local pair of low-metallicity, high-luminosity compact objects.
\end{abstract}


\section{Introduction} 

\indent Analysis of emission-line spectra provides the principal diagnostic tools for determining gas-phase physical characteristics of star-forming galaxies.
Heavy element abundances are valuable in the analysis of star formation histories (e.g., \citealp{bib:Kennicutt1998, bib:Contini2002}), chemical evolution (e.g., \citealp{bib:Steidel1996, bib:KobulnickyKewley2004, bib:Lamareille2006}), and general galactic evolution trends such as the luminosity-metallicity relation (e.g., \citealp{bib:Rubin1984, bib:Lee2004, bib:vanZee2006}), and the mass-metallicity relation (e.g., \citealp{bib:Lequeux1979, bib:Tremonti2004}).
Determination of this gas-phase property remains a difficult task for all but the most nearby and luminous \HII\ regions and emission-line galaxies (ELGs), and is intrinsically limited both by our ability to accurately measure intensity ratios of nebular emission lines and by our calibration of the conversion of observational data into abundance measurements.\\
\indent Metallicities determined using the temperature of the electron gas (the so-called direct- or \Te-method), considered to be the most accurate and robust available, are calculated using measurements of often exceedingly weak temperature-sensitive auroral emission lines such as \OIII$\lambda$4363 \citep{bib:OsterbrockFerland2006}. 
In computing the chemical composition of photoionized nebulae via the direct-method, the limiting factor is typically the measurement precision of the electron gas temperature.
Furthermore, the emission strength of the temperature-sensitive auroral lines scales with increasing \Te, and are thus preferentially detected in the hotter nebulae of low-abundance \HII\ regions and ELGs.
Determination of the electron temperature for high-metallicity systems is therefore a difficult observational task.
Reliable observations of high quality are thus relatively rare in the literature.\\
\indent For the majority of cases in which direct metallicity determinations are not possible, relations utilizing consistently detectable strong lines have been developed.
These are either empirical relations that are tied to direct-method abundances, theoretical relations that are based on photoionization models, or semi-empirical relations that are a mixture of the two (e.g., \citealp{bib:Moustakas2010}).
The most common such strong-emission-line (SEL) method is \R\ = (\OII$\lambda\lambda$3726,3729+\OIII$\lambda\lambda$4959,5007)/\HB, first introduced by \citet{bib:Pagel1979}.
\R\ provides an estimate of total cooling and thus represents a proxy for global metallicity \citep{bib:KewleyDopita2002}, but is famously double-valued at high- and low-abundance and includes an ambiguous turnaround region.
Calibration of this relation relies upon comparison to accurate measurements of metallicity, derived via the direct-method or through photoionization models (e.g., \citealp{bib:Kewley2001}).
Many \R\ calibrations of the latter sort are available in the literature (e.g., \citealp{bib:EdmundsPagel1984, bib:Skillman1989, bib:Zaritsky1994}).
As \Te-method abundances are more readily determined for lower-abundance systems, the \R\ relation is much better calibrated using direct-method data below the turnaround region than above \citep{bib:Pilyugin2000, bib:PilyuginThuan2005}.\\
\indent The primary goal of this study is to attempt to derive direct-method (\Te) abundances of star-forming galaxies for use in constraining the upper-metallicity branch of the \R\ relation.
We utilize the superior light-gathering capability of the Keck 10-m telescope to observe a specially selected set of strong-lined ELGs from the KISS sample \citep{bib:Salzer2000, bib:Salzer2001}.
These observations were planned with the hope of detecting and measuring the temperature-sensitive \OIII$\lambda4363$ emission line in higher-metallicity star-forming systems situated above the \R\ relation turnaround region.
With a robust calibration of \R, it becomes possible for the full range of ELG metallicities to be reliably studied, rather than being limited only to the brightest and most metal-poor objects.
Accurate abundances for high-metallicity targets allows for the comprehensive understanding of a more representative sample of star-forming galaxies in the local Universe.
Furthermore, the refinement of reliable empirical metallicity estimation methods from our observations of the total integrated light of the ELGs is particularly useful in the study of higher-redshift galaxies, where spectral observations are global by necessity and the detection of auroral lines becomes observationally difficult.
Calibration of the \R\ method using global spectra of local star-forming galaxies will provide for a more consistent approach to metallicity relations and chemical enrichment studies across cosmic time.\\
\indent There are often offsets in metallicity estimates when employing direct- and SEL-methods, where in general \Te-method abundances are low as compared to techniques utilizing ratios of strong lines (e.g., \citealp{bib:Kennicutt2003, bib:Bresolin2005, bib:ZuritaBresolin2012}).
Biases in the sample selection of observational targets used to calibrate some empirical relations can be a major contributor to such abundance discrepancies.
The magnitude of these offsets is typically of order a few tenths of a dex.
A more comprehensively populated \R\ relation, including \Te-method abundances of the upper-metallicity branch, will help to reduce such discrepancies.
An alternative approach to resolving this disparity involves a reexamination of the \Te-method.
In computing direct-method metallicities, the electron temperature can be estimated from ions of similar electronic structure to O$^{++}$, such as doubly-ionized sulfur (e.g., \citealp{bib:Garnett1989, bib:Bresolin2005, bib:Perez-Montero2006}).
Models show that the S$^{++}$ zone straddles the zones of both O$^{+}$ and O$^{++}$ \citep{bib:Garnett1992}.
The electron temperature derived from \SIII\ emission lines may therefore be a better \emph{single} representation of the temperature of the nebula as a whole.\\
\indent In \S2 we discuss the galaxy sample, the observations, and the data-reduction methods.
In \S3 we present the spectral data obtained using the W. M. Keck Observatory.
Section 4 presents a detailed analysis of the 13 objects with abundance-quality spectra.
We measure properties of the nebular emission regions such as electron temperature and electron density and calculate abundance ratios of heavy elements with respect to hydrogen.
Section 5 discusses the abundance results, presenting the preliminary findings of our program to constrain the upper-metallicity branch of the \R\ empirical strong-line relation.
We do so with direct (\Te) method abundances of relatively metal-rich ELGs, using the KISS sample for target selection and the Keck telescope to deliver high-quality spectra in order to detect weak temperature-sensitive emission lines.
In addition, we compare our abundance results with various strong-emission-line (SEL) abundance determination methods and calibrations available in the literature, and compare metallicity estimates computed via the direct method using electron temperatures derived from both doubly-ionized oxygen (O$^{++}$) and sulfur (S$^{++}$) to test the effect of sampling different zones of ionization within the nebula.
Finally, we scrutinize our sample to better understand the physical limitations inherent to studies of this nature, compare direct-method abundances calculated using electron temperatures characteristic of different ionization zones, and explore in more detail the three categories of objects uncovered in our observations.
Our results are summarized in \S6.

\section{Observations and Data Reduction} 

\subsection{Sample Selection} 

\begin{figure*}
\plotone{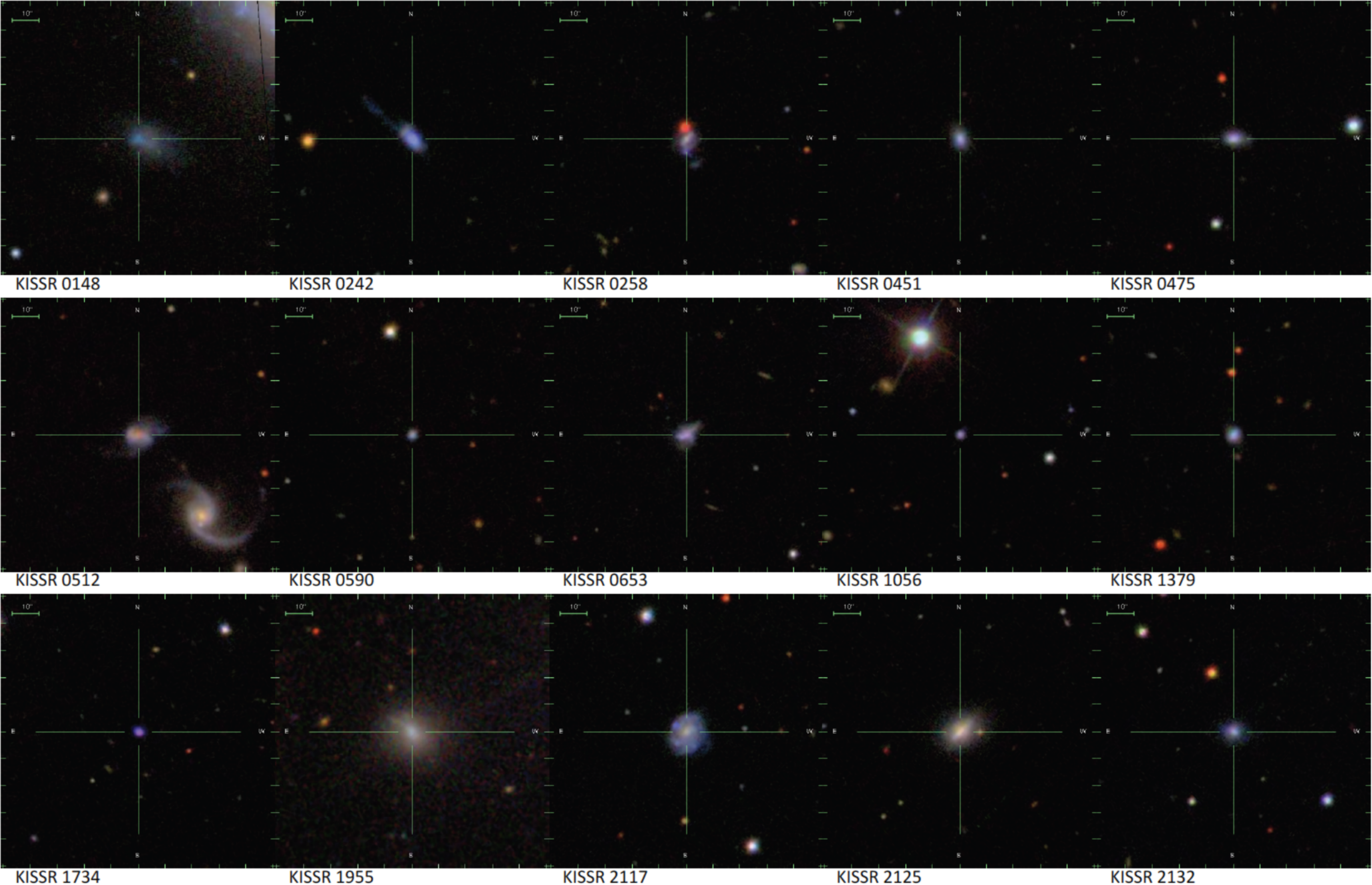}
\vspace{0.2cm}
\caption{The fifteen galaxies selected from KISS observed for this study, as taken from the Sloan Digital Sky Survey (SDSS) \citep{bib:Abazajian2005}.
Targets range in redshift between $z$ = 0.00791 and $z$ = 0.09115.
Image boxes are 100 arcsec per side and in some cases the objects are seen to be quite compact.}
\label{fig:SDSS_images_P}
\vspace{-0.5cm}
\end{figure*}

\begin{figure*}
\plotone{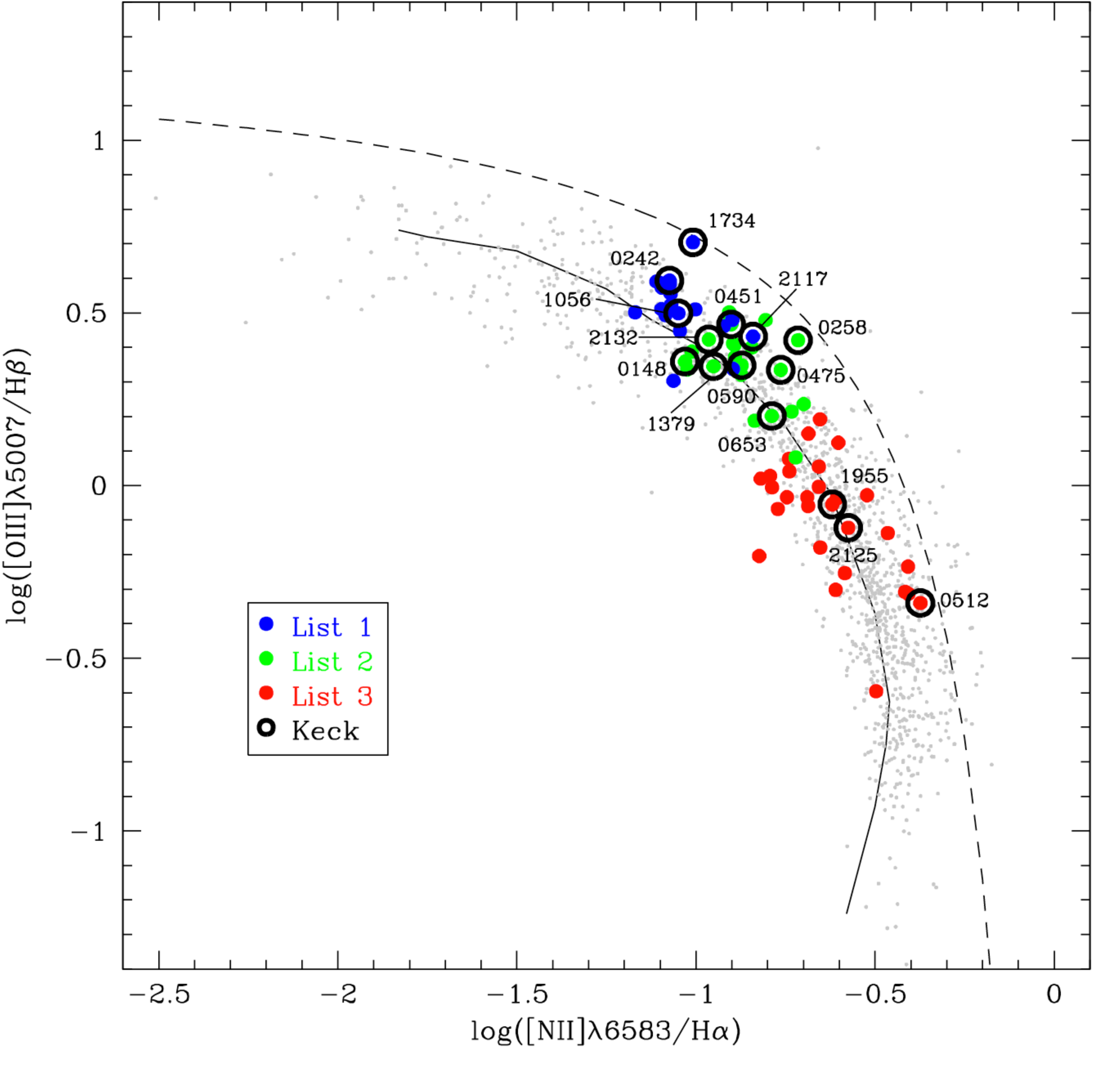}
\vspace{-0.5cm}
\caption{Diagnostic diagram of KISS star-forming galaxies using emission-line ratios from the original survey data, with galaxies from each of the three lists of varying selection criterion overlaid as blue circles (List 1: High \OIII\ EWs), green circles (List 2: Lower \OIII\ EWs), and red circles (List 3: Higher Coarse Abundances).
The fifteen galaxies to be observed with Keck for this study are labeled and signified with black rings.
The solid line represents a star formation sequence derived from theoretical models \citep{bib:DopitaEvans1986}, while the dashed line represents an empirically defined demarcation between starburst galaxies and AGN \citep{bib:Kauffmann2003}.}
\label{fig:KoK_DD}
\vspace{0.5cm}
\end{figure*}

\indent Our spectroscopic targets for this project were potentially metal-rich galaxies identified by the KPNO International Spectroscopic Survey (KISS; \citealp{bib:Salzer2000}).
KISS used low-dispersion objective-prism spectra to identify ELG candidates via detection of line emission in galaxies with redshifts of less than 0.095.
The ``red" survey \citep{bib:Salzer2001, bib:Gronwall2004b, bib:Jangren2005} cataloged objects by means of the \HA\ line, while the ``blue" survey \citep{bib:Salzer2002} distinguished objects through strong \OIII$\lambda$5007 line emission.
The KISS sample is comprised of a wide variety of ELGs, including starburst nucleus galaxies, \HII\ galaxies, irregular galaxies with significant star formation, and blue compact dwarfs (BCDs).
In addition, the survey detected low-ionization nuclear emission-line regions (LINERs) and Seyfert galaxies as well as quasars whose redshifts place their emission lines into the objective-prism spectral bandpass.\\ 
\indent The galaxies discussed in this paper were chosen from the KISS \HA-selected catalog with selection criteria based on a combination of \OIII\ line strength and coarse abundance estimates \citep{bib:MelbourneSalzer2002, bib:Salzer2005b}.
We were specifically interested in objects for which \OIII$\lambda$4363 might be detectable in order to obtain the nebular electron temperature \Te, which is measured using the line ratio \OIII$\lambda$4363/\OIII$\lambda\lambda$4959,5007 (e.g., \citealp{bib:Izotov1994}).
Additionally, we are interested in galaxies with high enough oxygen abundance to be located above the turnaround region of the \R\ relation (\abun\ $\gtrsim$ 8.3) as a means to constrain/calibrate the location of the upper-metallicity branch.\\
\indent The selection of targets for this study was based on the emission-line data obtained from ``quick-look" follow-up spectra of KISS galaxies \citep{bib:Wegner2003, bib:Melbourne2004, bib:Gronwall2004a, bib:Jangren2005, bib:Salzer2005a}.
Based on the published spectral data, we were able to generate lists of potentially interesting objects.
These spectra were usually not of high enough quality for the measurement of direct abundances, but they did yield reasonably accurate strong-line ratios that gave a good estimate of the properties of each object.\\
\indent Galaxies were chosen for observation based on their inclusion in one of three lists of KISS objects generated with varying selection criteria.
First, our targets were selected to be among those with the highest line strengths, in order to have the best chance of detecting the weak temperature-sensitive emission lines.
Using the equivalent width (EW) of the \OIII$\lambda$5007 line measured in our quick-look spectra as a discriminant, list 1 requires an EW \textgreater\ 100 \AA, list 2 requires an EW between 50 and 100 \AA, and list 3 requires an EW between 5 and 50 \AA.
Second, our objects require higher abundances, as our goal is to probe into the upper-metallicity branch of the \R\ relation.
Using coarse abundance estimates \citep{bib:MelbourneSalzer2002, bib:Salzer2005b}, list 1 requires a \abun\ \textgreater\ 8.2, list 2 requires a \abun\ \textgreater\ 8.3, and list 3 requires a \abun\ \textgreater\ 8.5.
For the third list, additional selection criteria include the requirement that the Balmer decrement \cHB\ \textless\ 0.7 in order to avoid the most heavily reddened objects.
The number of galaxies selected to each of the three lists is as follows:
List 1 includes 20 objects, list 2 includes 22 objects, and list 3 includes 28 objects.
In total, four objects were observed from list 1, eight objects from list 2, and three objects from list 3, for fifteen total objects observed during our run.
Images of each of the fifteen galaxies taken from the Sloan Digital Sky Survey (SDSS; \citealp{bib:Abazajian2005}) are shown as Figure \ref{fig:SDSS_images_P}.\\
\indent The galaxies selected for this study are displayed on a spectral activity diagnostic diagram (e.g., \citealp{bib:Baldwin1981}) in Figure \ref{fig:KoK_DD} using emission-line ratios taken from the original KISS survey data, overlaying data points representing all star-forming galaxies from KISS.
Objects from the three target lists are indicated by symbols of different color, while those to be observed for this study with Keck are further indicated by the heavy black rings.
The solid line represents a star-formation sequence derived by \citet{bib:DopitaEvans1986} from theoretical models, while the dashed line represents an empirically defined demarcation between starburst galaxies and AGN from \citet{bib:Kauffmann2003}.
It should be noted that all initial observations of relevant line ratios place the galaxies considered for this study firmly within the boundaries of regular star-forming galaxies (i.e., no targets were believed to exhibit spectral contamination from an AGN).
In general, these criteria allowed us to identify starburst galaxy candidates of medium- to high-abundance with a moderate degree of success.
In a few cases, however, we discovered examples of galaxies with spectral contamination from AGN and galaxies with anomalously low abundances and high luminosities for their redshifts.
We will explore the inhomogeneity of these targets within our sample in greater detail in \S5.2.
A list of all observed galaxies with coordinates and basic information is presented as Table 1, including references to important information relevant to KISS from these papers.
\begin{deluxetable*}{ccccccccccc}
\tablenum{1}
\tabletypesize{\scriptsize}
\tablewidth{0pt}
\tablecaption{Object Parameters and Identifiers for Observed Galaxies}
\tablehead{\colhead{KISSR}&\colhead{Field}&\colhead{KISS ID}&\colhead{RA}&\colhead{Dec}&\colhead{$m_{B}$}&\colhead{$M_{B}$}&\colhead{$z$}&\colhead{$t_{\text{blue}}$ [sec]}&\colhead{$t_{\text{red}}$ [sec]}&\colhead{$t_{\text{NIR}}$ [sec]}}
\startdata
0148 & F1250 & 317 & 12:54:45.4 & +28:55:31 & 17.06 & -15.93 & 0.00791 & 1$\times$600 & 1$\times$600 & --- \\
0242 & F1315 & 3582 & 13:16:03.9 & +29:22:54 & 16.71 & -19.46 & 0.03782 & 2$\times$900, 1$\times$300 & 2$\times$900 & 1$\times$300 \\
0258 & F1315 & 1234 & 13:18:40.3 & +28:59:57 & 17.57 & -20.15 & 0.07567 & 2$\times$600 & 2$\times$600 & --- \\
0451 & F1405 & 3370 & 14:06:35.0 & +29:27:37 & 17.81 & -18.59 & 0.04158 & 3$\times$600 & 2$\times$600 & 1$\times$600 \\
0475 & F1410 & 4626 & 14:11:06.6 & +29:22:21 & 17.58 & -19.89 & 0.06641 & 3$\times$600 & 2$\times$600 & 1$\times$600 \\
0512 & F1415 & 2664 & 14:17:48.4 & +29:03:11 & 17.36 & -20.46 & 0.07719 & 2$\times$900, 1$\times$600 & 2$\times$900 & 1$\times$600 \\
0590 & F1455 & 8836 & 14:55:01.2 & +29:33:31 & 19.28 & -16.48 & 0.03072 & 3$\times$600 & 2$\times$600 & 1$\times$600 \\
0653 & F1510 & 4675 & 15:10:52.7 & +28:46:46 & 17.53 & -20.10 & 0.06990 & 3$\times$600 & 2$\times$600 & 1$\times$600 \\
1056 & F1635 & 4097 & 16:37:18.2 & +29:51:11 & 19.32 & -18.91 & 0.09115 & 2$\times$600 & 2$\times$600 & --- \\
1379 & F1246 & 1011 & 12:48:13.4 & +43:57:04 & 18.02 & -17.67 & 0.02969 & 3$\times$600 & 2$\times$600 & 1$\times$600 \\
1734 & F1411 & 4039 & 14:10:59.2 & +43:02:47 & 18.44 & -18.97 & 0.06556 & 2$\times$600, 1$\times$300 & 2$\times$600 & 1$\times$300 \\
1955 & F1528 & 4743 & 15:27:59.4 & +42:50:58 & 16.83 & -17.88 & 0.01849 & 1$\times$900 & 1$\times$900 & --- \\
2117 & F1610 & 7310 & 16:09:53.6 & +44:13:22 & 16.58 & -19.59 & 0.03716 & 3$\times$600 & 2$\times$600 & 1$\times$600 \\
2125 & F1610 & 5338 & 16:10:20.4 & +43:00:35 & 16.58 & -18.73 & 0.02517 & 1$\times$1200, 1$\times$300 & 1$\times$1200 & 1$\times$300 \\
2132 & F1610 & 3202 & 16:11:45.8 & +43:30:44 & 17.78 & -18.37 & 0.03685 & 2$\times$600 & 2$\times$600 & --- \\
\enddata
\label{tab:ObjParam}
\end{deluxetable*}

\subsection{Observations} 

\begin{figure*}
\epsscale{0.736}
\plotone{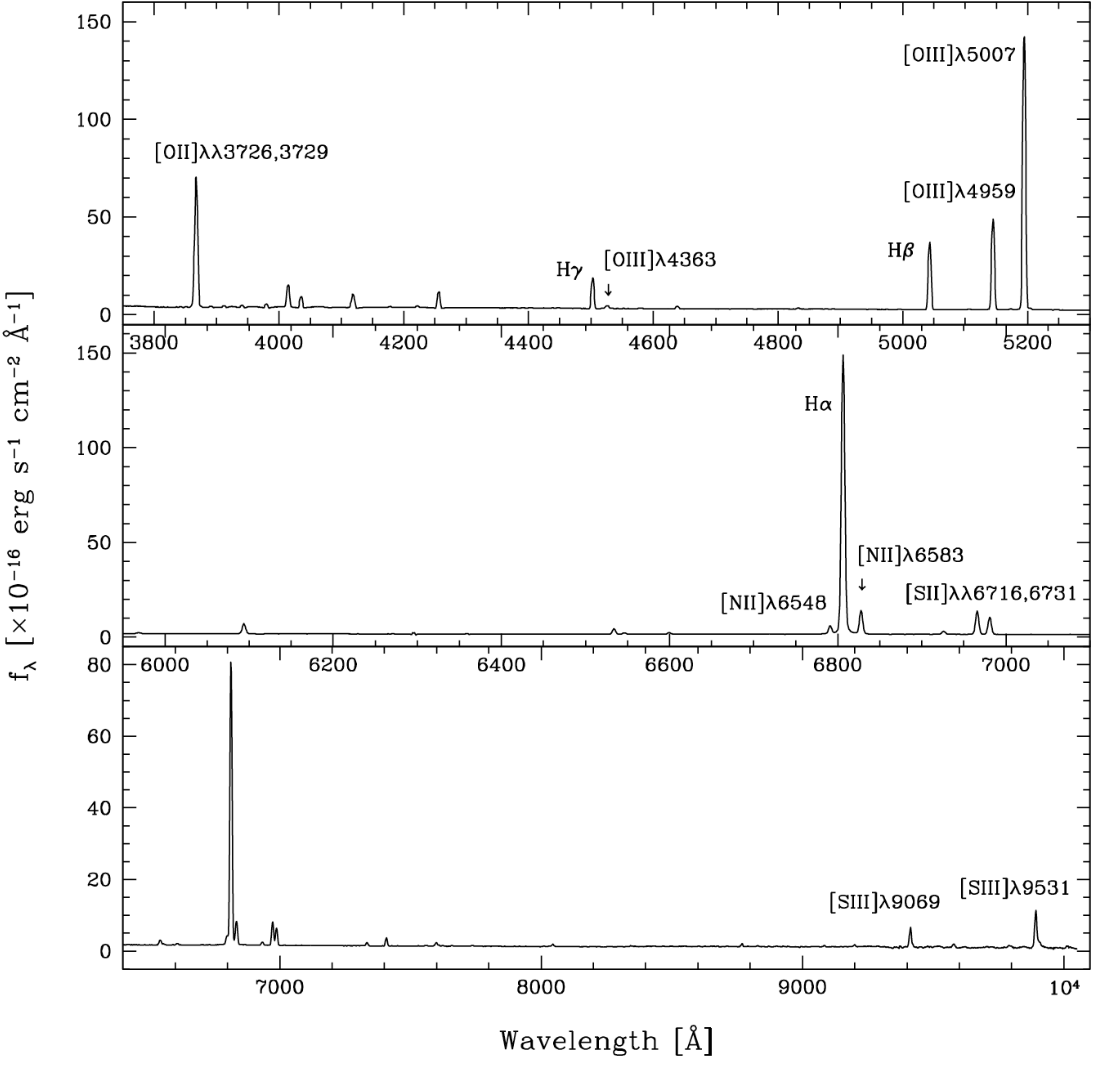}
\vspace{-0.3cm}
\caption{Optical spectrum of KISSR 0242, in all three wavelength regimes afforded by Keck LRIS, including ``blue", ``red", and ``infrared".
Emission lines that are important to abundance analysis for this study are labeled.}
\label{fig:kr0242labelP}
\end{figure*}

\begin{figure*}
\epsscale{0.736}
\plotone{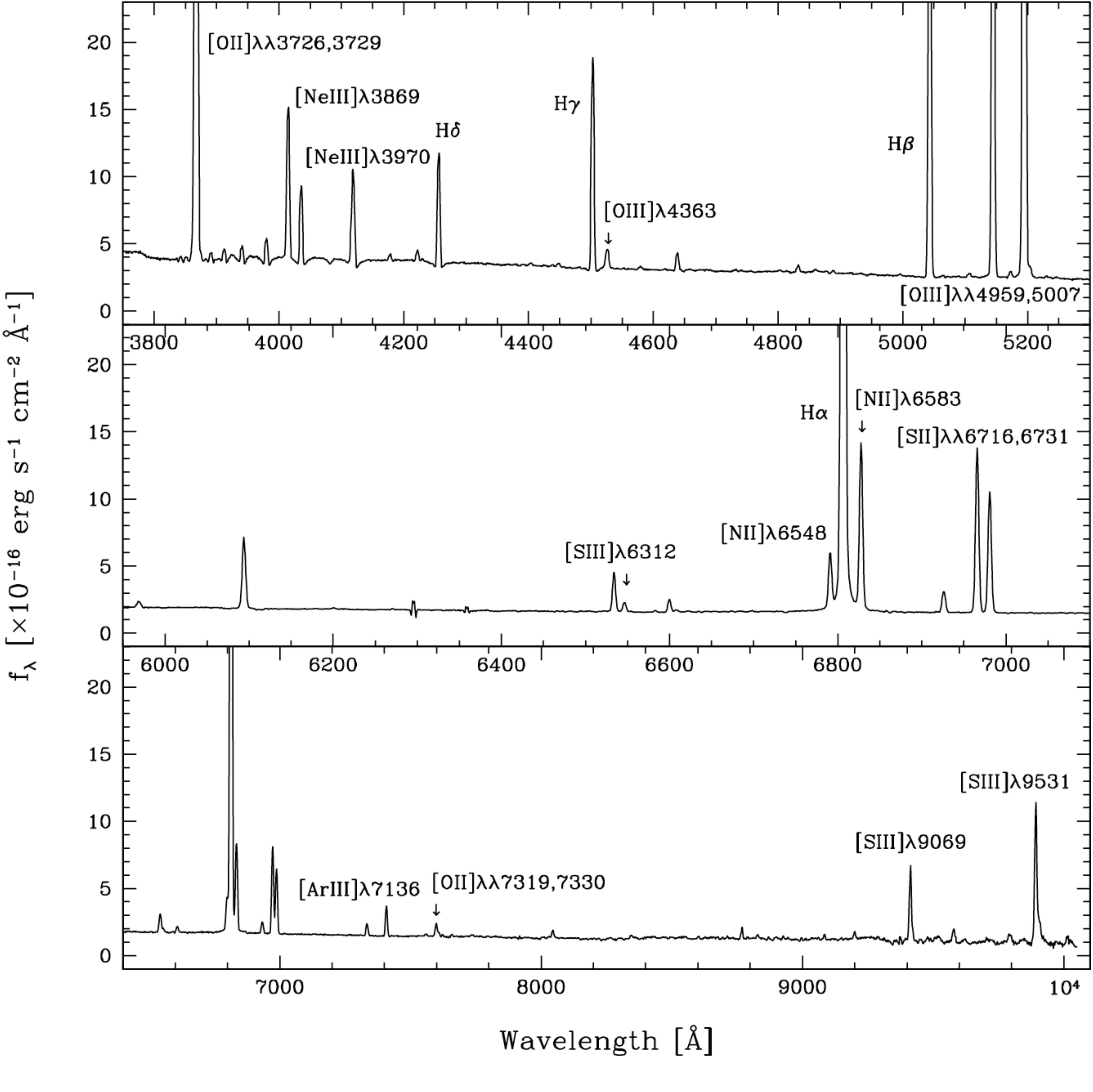}
\vspace{-0.3cm}
\caption{Same as Figure \ref{fig:kr0242labelP}, with reduced vertical scaling to highlight emission line detail.
Emission lines important for the abundance analysis for this study are labeled.}
\label{fig:kr0242biglabelP}
\end{figure*}

\indent Spectra of 15 KISS ELGs were obtained using the Keck I 10 m telescope on 25 May 2006 with the Low-Resolution Imaging Spectrometer (LRIS), a double spectrograph that includes a dichroic that directs light toward a red and a blue side \citep{bib:Oke1995}.
The use of two distinct dispersion elements and cameras enables simultaneous spectral coverage spanning from $\sim$3000 \AA\ to $\sim$10,000 \AA\ with good dispersion.
The KISS catalog is comprised of galaxies characterized by a single (often central) dominant unresolved source of emission, which for our targets were mostly encompassed within our 1".5 wide slit.
For each observation, the slit was aligned with the parallactic angle to accommodate for differential atmospheric dispersion.
Blue side exposures used the 600/400 grism, with a dispersion of 0.63 \AA\ pixel$^{-1}$, a resolution of $\sim$6 \AA, and wavelength coverage between $\sim$3000 \AA\ and $\sim$5600 \AA.
Red side exposures used the 900/5500 grating, with a dispersion of 0.53 \AA\ pixel$^{-1}$, a resolution of $\sim$3 \AA, and wavelength coverage between $\sim$5700 \AA\ and $\sim$7300 \AA.
Each galaxy was first observed using the blue and red sides simultaneously to determine if further observations would yield usable spectra.
If the raw spectra were of substantially high quality, the red side grating was switched to observe NIR lines.
This was completed for ten of the fifteen galaxies.
For NIR exposures, the red side was switched to the 400/8500 grating, with a dispersion of 1.16 \AA\ pixel$^{-1}$, a resolution of $\sim$6 \AA, and wavelength coverage between $\sim$6300 \AA\ to $\sim$10,000 \AA\ in order to observe the near-infrared \SIII\ nebular lines at $\lambda$9069 and $\lambda$9531.
Typical individual exposure times were 600 seconds, although some targets necessitated individual exposure times of 1200, 900, or 300 seconds.
Exposure time information is listed in Table 1.\\
%
\indent Example spectra of one galaxy from this study (KISSR 0242) seen as Figures \ref{fig:kr0242labelP} and \ref{fig:kr0242biglabelP} illustrates the wavelength coverage afforded by Keck LRIS, extending from the near-UV to the near-IR.
Furthermore, Figure \ref{fig:kr0242biglabelP} reveals clear detection of, e.g., the \OIII$\lambda$4363 line required to derive \Te-method oxygen abundances.
This object is of particularly high excitation and signal strength and should not be taken as typical for our study, but rather its spectral characteristics were chosen to demonstrate the capabilities of our apparatus and observing program.
For five of the fifteen galaxies in our sample, an inspection of the first two exposures indicated that the object did not possess adequately strong emission lines to warrant further observation, and so the red side grating was not switched and thus observations of the near-infrared \SIII\ lines were not acquired.
Our observations included spectra of a Hg-Cd-Zn and Ar-Ne lamps to set the wavelength scale as well as observations of the spectrophotometric standard stars GD 153 \citep{bib:Bohlin1995}, Feige 56 \citep{bib:Hamuy1992}, Feige 66, BD+28 4211, and Feige 110 \citep{bib:Massey1988} for flux calibration.

\subsection{Data Reduction} 

\indent The data reduction was carried out with the Image Reduction and Analysis Facility (\texttt{IRAF}\footnote{\texttt{IRAF} is the Image Reduction and Analysis Facility distributed by the National Optical Astronomy Observatory, which is operated by the Association of Universities for Research in Astronomy (AURA) under cooperative agreement with the National Science Foundation (NSF).}).
Processing of the two-dimensional spectral images followed standard methods.
All of the reduction steps mentioned below were carried out on the red and blue spectral images independently.
The mean bias level was determined and subtracted automatically from each image by the data acquisition software.
A mean bias image was created by combining 11 zero-second exposures taken on the night of the observations.
This image was subtracted to correct the science images for any possible two-dimensional structure in the bias.
Flat-fielding was achieved using an average-combined quartz lamp image that was corrected for the wavelength-dependent response of the system.
For cosmic ray rejection, we used \texttt{L.A.Cosmic} \citep{bib:vanDokkum2001}.\\
\indent One-dimensional spectra were extracted using the \texttt{IRAF} \texttt{APALL} routine.
The extraction width (i.e., distance along the slit) was selected independently for each source by examination.
As the LRIS spectrograph uses a dichroic beam splitter to break the spectrum into red and blue sections, in order to analyze the data it is necessary that the two sections are on a consistent flux scale with each other, accomplished by setting the extraction regions to be the same.
For compact targets our observations acquired global spectra, while for more extended galaxies we analyzed the nuclear region only.
In all cases, the extraction width was limited to 10 pixels or 2.15 arcseconds for the red side CCD (pixel scale of 0.215'' pix$^{-1}$).
For the blue side CCD we scaled the extraction width accordingly to 15.62 pixels (from a pixel scale of 0.135" pix$^{-1}$).  
Sky subtraction was also performed at this stage, with the sky spectrum being measured in regions on either side of the object extraction window.
The Hg-Cd-Zn and Ar-Ne lamp spectra were used to assign a wavelength scale, and the spectra of the spectrophotometric standard stars were used to establish the flux scale.
The standard star data were also used to correct the spectra of our target ELGs for telluric absorption.
This is important because at the redshifts of some of our targets the \SII\ doublet falls in the strong telluric $B$-band and would lead to a severe underestimate of the true emission-line flux.\\
\indent We utilized the galaxy observations to measure any low-level flux scale differences between the red and blue sides.
To do this, we measured the continuum flux of the galaxies in 50 \AA\ bins from 5200 \AA\ to 6000 \AA.
A plot of flux versus wavelength reveals any small shifts between the continuum fluxes on either side of the dichroic break.
Since the spectral extraction regions covered the same angular extent on both sides, the correction factors were expected to be small.
This turned out to be the case, with the derived correction factors for each observation averaging a few percent.
For our instrumental setup, \HA\ and \HB\ are always on opposite sides of the dichroic.
Therefore, the main effect of the correction factor is to modify the \HA/\HB\ ratio and the value of the reddening parameter \cHB.
This necessary procedure introduces an additional level of uncertainty into the derivation of the nebular abundances in the sense that it affects reddening corrections.
Great care was taken in ensuring that the red and blue sections of the spectra were extracted consistently.

\section{The Spectral Data} 

\begin{figure*}
\plotone{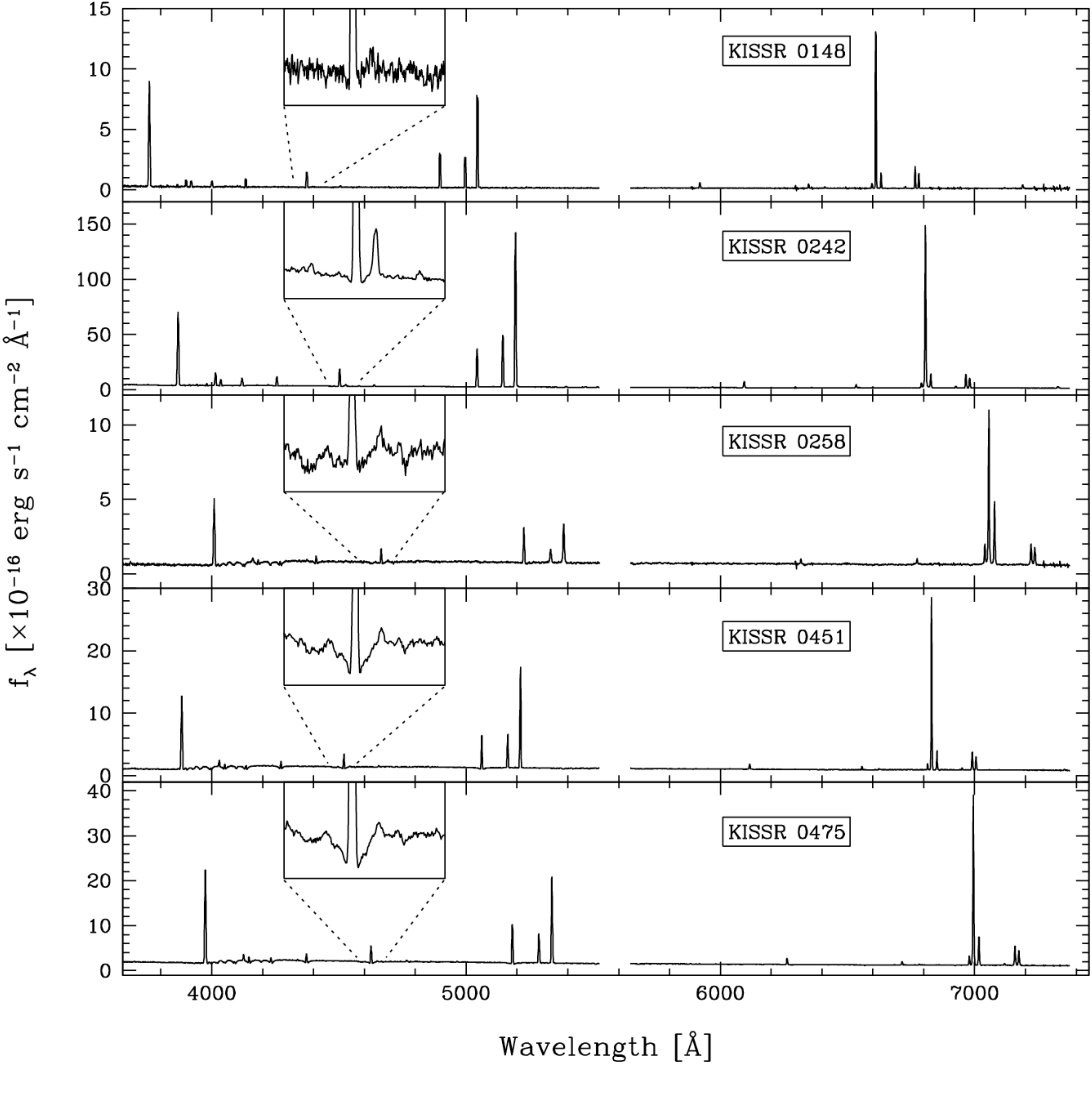}
\caption{Spectra of ``blue" and ``red" wavelength ranges of the first five galaxies studied for this project, with inset showing the \HG\ and \OIII$\lambda$4363 emission lines.}
\vspace{-0.7cm}
\label {fig:block1}
\end{figure*}

\begin{figure*}
\plotone{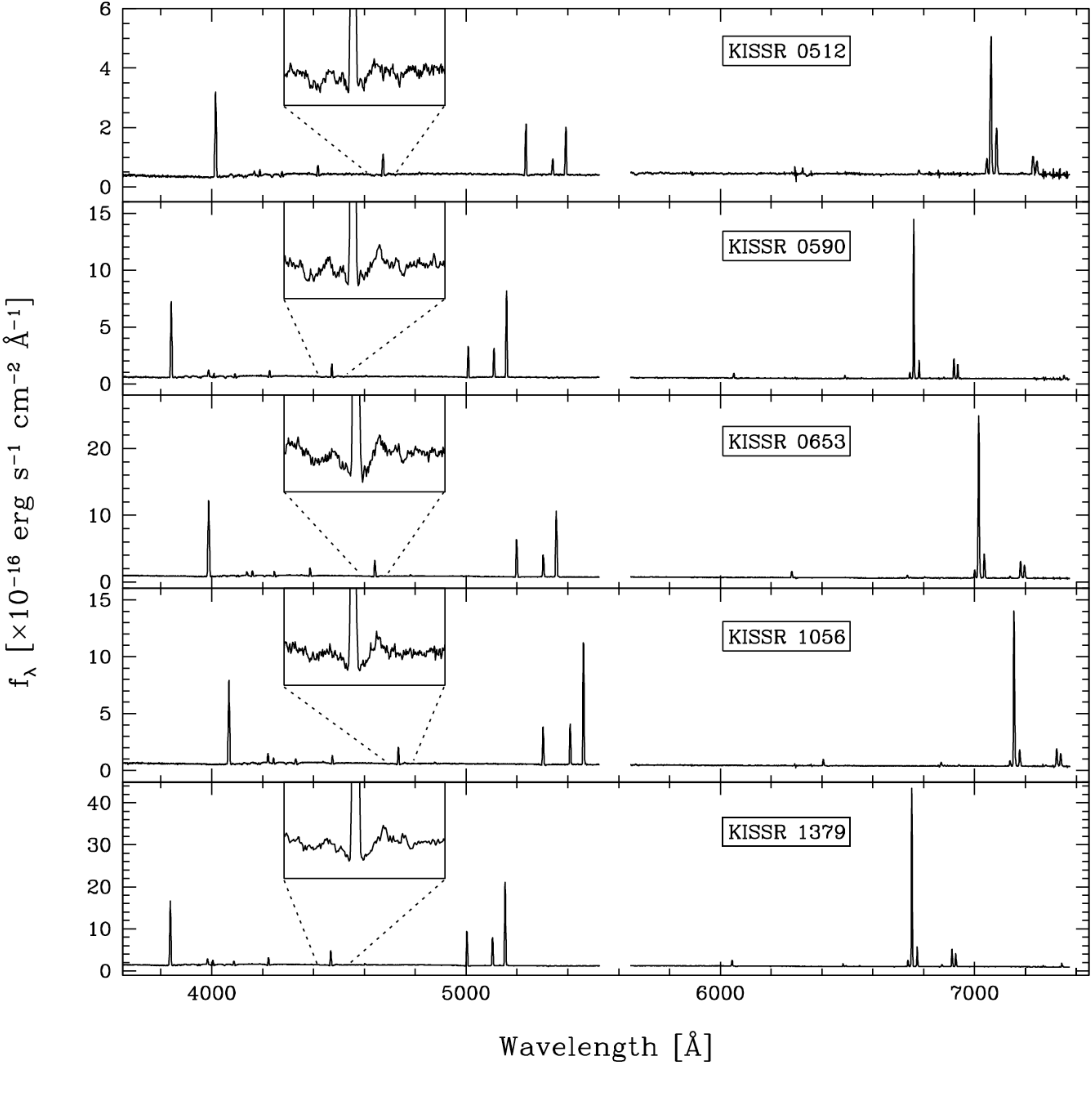}
\caption{Spectra of ``blue" and ``red" wavelength ranges of the next five galaxies studied for this project, with inset showing the \HG\ and \OIII$\lambda$4363 emission lines.}
\vspace{-0.7cm}
\label{fig:block2}
\end{figure*}

\begin{figure*}
\plotone{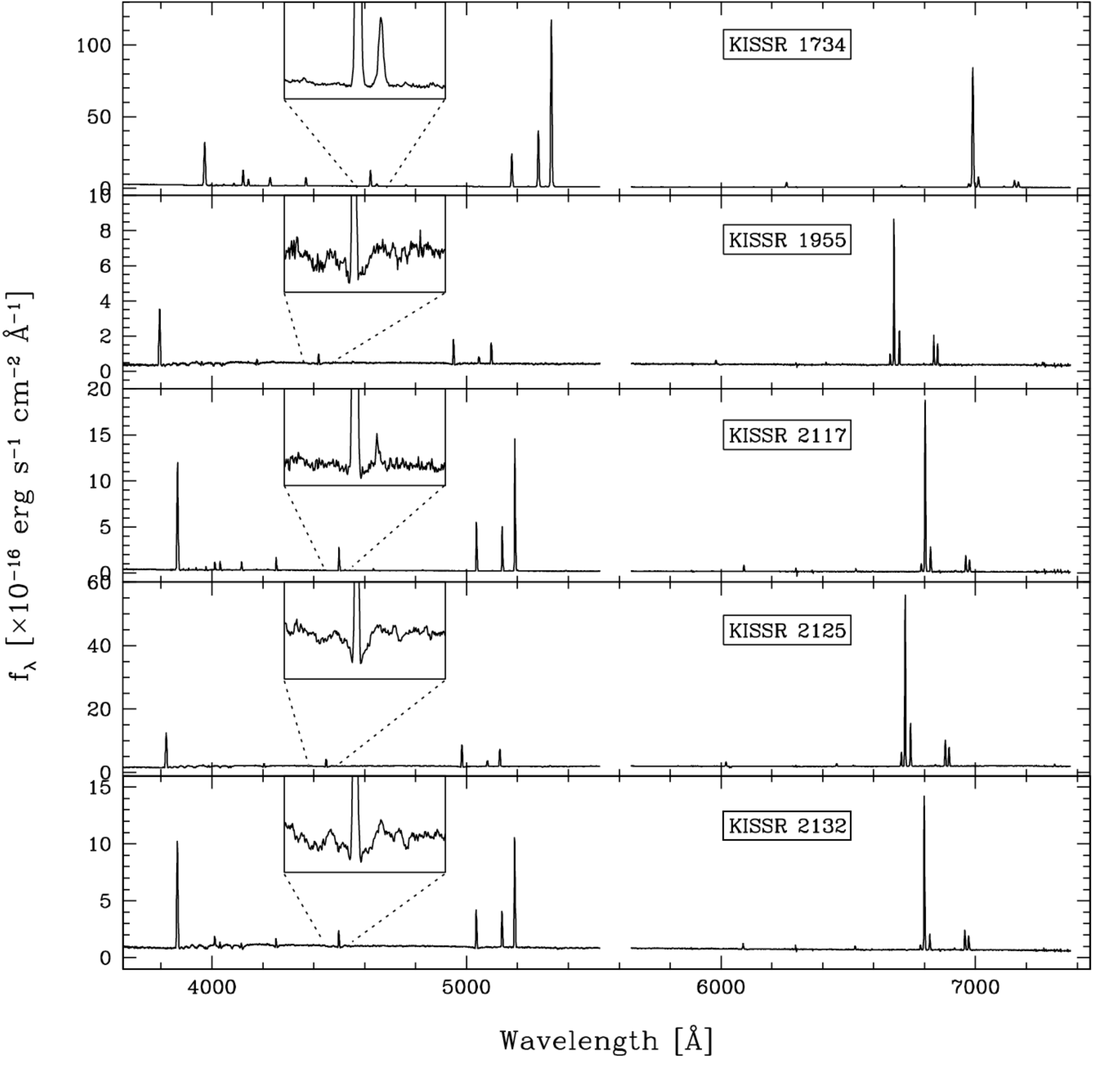}
\caption{Spectra of ``blue" and ``red" wavelength ranges of the last five galaxies studied for this project, with inset showing the \HG\ and \OIII$\lambda$4363 emission lines.}
\vspace{-0.7cm}
\label{fig:block3}
\end{figure*}

\indent Emission-line flux measurements were carried out using the \texttt{IRAF} routine \texttt{SPLOT}.
When multiple images of the same object were available in any of the three wavelength ranges, the spectra were reduced separately then combined into a single high S/N spectrum prior to the measurement stage.
For an initial estimate of the internal reddening in each galaxy we calculate \cHB\ from the \HA/\HB\ line ratio.
The value of \cHB\ is then used to correct the measured line ratios for reddening, following the standard procedure (e.g., \citealp{bib:OsterbrockFerland2006}),
\[
\frac{I(\lambda)}{I(\text{H}\beta)} = \frac{F(\lambda)}{F(\text{H}\beta)}\ 10^{[c_{\text{H}\beta} f(\lambda)]},
\]
\noindent where $f(\lambda$) is derived from studies of absorption in the Milky Way (using values taken from \citealp{bib:Rayo1982}).
The final \cHB\ values used for the reddening correction were determined from a simultaneous fit to the reddening and underlying stellar absorption in the Balmer lines, using all available Balmer line ratios.
Under the simplifying assumption that the same EW of underlying absorption applies to each Balmer line, an absorption line correction was applied to each spectrum that ranged in value from 0 to 5 \AA.
The value for the underlying absorption was varied until a self-consistent value of \cHB\ for the Balmer line ratios, \HA/\HB, \HG/\HB, and \HD/\HB, was found.
This process led to the determination of a characteristic value for the underlying absorption for each galaxy.
The average value was typically equal to $\sim$2 \AA\ (e.g., \citealp{bib:McCall1985, bib:SkillmanKennicutt1993}); line strengths of the Balmer emission lines were adjusted accordingly.
In the case of KISSR 0512, the derived value for \cHB\ is slightly negative, although the formal error in \cHB\ is such that the measured value is consistent with \cHB\ = 0.00.
We therefore adopt \cHB\ = 0.00 for the computation of reddening-corrected line ratios for this object.
Results of our spectroscopic observations of the Keck sample, presented as reddening-corrected line ratios relative to \HB, are presented in Table 2.\\
\indent Determinations of \cHB\ for each galaxy derived from the earlier follow-up spectroscopy of the original KISS galaxies  were used as a source of comparison for our study \citep{bib:Wegner2003, bib:Melbourne2004, bib:Gronwall2004a, bib:Jangren2005, bib:Salzer2005a}.
In the majority of cases, these values agree.
For some galaxies, however, disparities in value may represent cases in which the placement of the slit was different between the two sets of observations, and as such properties of different regions of the galaxy were measured (e.g., a nuclear emission region compared to a knot of \HII\ in a spiral arm).
The spectra of our fifteen objects are presented in Figures \ref{fig:block1}, \ref{fig:block2}, and \ref{fig:block3}.

\section{Metal Abundances} 

\subsection{Electron Density and Temperature} 

\indent In determining the physical characteristics of \HII\ regions and ELGs, a two-zone model for the star-forming nebulae is typically assumed.
The outer zone is represented by a lower-temperature region in which oxygen is singly-ionized, while the inner zone is a medium-temperature zone in which oxygen is doubly-ionized.
Hydrogen is assumed to be fully ionized within the radius of the outer zone, and is assumed to be neutral outside this zone.\\
\indent In each zone we would ideally like to know both the electron density and temperature.
The available observational diagnostics, however, only allow for an electron density measurement in the low-temperature zone and an electron temperature measurement in the medium-temperature zone.
We use the \SII\ $\lambda$6716/$\lambda$6731 \citep{bib:OsterbrockFerland2006} ratio to determine the electron density, which in all measurable cases is roughly 100 e$^{-}$ cm$^{-3}$.
For galaxies for which electron density was not derived from our observations, we assume a density of 100 e$^{-}$ cm$^{-3}$.
Our abundance analysis assumes that the density does not vary from zone to zone.

\subsubsection{Oxygen \Te} 

\indent The electron temperature in the medium-temperature zone is determined by the oxygen line ratio \OIII$\lambda$4363/\OIII$\lambda\lambda$4959,5007, which represents a comparison in doubly-ionized oxygen of electron transitions with considerably different excitation energies \citep{bib:OsterbrockFerland2006}.
As can be seen by inspection of the insets in Figures \ref{fig:block1}, \ref{fig:block2}, and \ref{fig:block3}, several of our $\lambda$4363 detections are quite weak, and great care was required in their measurement.
The uncertainties in the final $\lambda$4363 line fluxes are reflected in the large uncertainties in the derived values of \Te.
Calculations of both the electron density and electron temperature are carried out using the program \texttt{ELSA} \citep{bib:Johnson2006}, which makes use of the latest collision strengths and radiative transition probabilities.\\
\indent We estimate the temperature in the low-temperature zone by using the algorithm presented in \citet{bib:Skillman1994} based on the nebular models of \citet{bib:Stasinska1990},
\[
t_{e}(\text{O}^{+}) = 2[t_{e}(\text{O}^{++})^{-1} + 0.8]^{-1},
\]
\noindent where $t_{e}$ are temperatures measured in units of 10$^{4}$ K.  The measured electron densities and temperatures are presented in Table 3.\\

\begin{turnpage}
\begin{deluxetable*}{cccccccccc}
\tabletypesize{\scriptsize}
\tablenum{2}
\tablewidth{0pt}
\tablecaption{Emission-Line Intensity Ratios Relative to \HB\ for Observed Galaxies (1 of 2)}
\tablehead{\colhead{Ion}&\colhead{$\lambda$ [\AA]}&\colhead{KISSR 0148}&\colhead{KISSR 0242}&\colhead{KISSR 0258}&\colhead{KISSR 0451}&\colhead{KISSR 0475}&\colhead{KISSR 0512}&\colhead{KISSR 0590}&\colhead{KISSR 0653}}
\startdata
\OII\ & 3727 & 3.287 $\pm$ 0.135 & 2.816 $\pm$ 0.115 & 3.241 $\pm$ 0.140 & 3.908 $\pm$ 0.164 & 3.582 $\pm$ 0.147 & 1.767 $\pm$ 0.039 & 3.501 $\pm$ 0.146 & 2.939 $\pm$ 0.122 \\
H 10 & 3798 & 0.042 $\pm$ 0.004 & 0.040 $\pm$ 0.002 & --- & --- & --- & --- & --- & 0.025 $\pm$ 0.002 \\
H 9 & 3835 & 0.065 $\pm$ 0.004 & 0.064 $\pm$ 0.002 & 0.064 $\pm$ 0.008 & --- & --- & --- & --- & 0.052 $\pm$ 0.003 \\
\NeIII\ & 3869 & 0.201 $\pm$ 0.008 & 0.404 $\pm$ 0.015 & 0.134 $\pm$ 0.009 & 0.243 $\pm$ 0.010 & 0.171 $\pm$ 0.007 & 0.050 $\pm$ 0.004 & 0.222 $\pm$ 0.010 & 0.106 $\pm$ 0.005 \\
He I+H 8 & 3889 & 0.175 $\pm$ 0.007 & 0.198 $\pm$ 0.007 & 0.140 $\pm$ 0.009 & 0.161 $\pm$ 0.007 & 0.168 $\pm$ 0.007 & 0.102 $\pm$ 0.005 & 0.152 $\pm$ 0.007 & 0.170 $\pm$ 0.007 \\
\NeIII+\HE\ & 3970 & 0.184 $\pm$ 0.007 & 0.271 $\pm$ 0.009 & 0.124 $\pm$ 0.008 & 0.133 $\pm$ 0.005 & 0.141 $\pm$ 0.005 & 0.090 $\pm$ 0.004 & 0.160 $\pm$ 0.007 & 0.136 $\pm$ 0.005 \\
He I & 4026 & --- & 0.013 $\pm$ 0.001 & --- & --- & --- & --- & --- & --- \\
\SII\ B & 4073 & --- & 0.023 $\pm$ 0.001 & 0.045 $\pm$ 0.007 & 0.036 $\pm$ 0.002 & 0.027 $\pm$ 0.002 & 0.034 $\pm$ 0.004 & 0.032 $\pm$ 0.003 & 0.024 $\pm$ 0.002 \\
\HD\ & 4102 & 0.249 $\pm$ 0.009 & 0.272 $\pm$ 0.008 & 0.234 $\pm$ 0.010 & 0.235 $\pm$ 0.008 & 0.240 $\pm$ 0.008 & 0.194 $\pm$ 0.006 & 0.239 $\pm$ 0.009 & 0.236 $\pm$ 0.008 \\
\HG\ & 4340 & 0.456 $\pm$ 0.013 & 0.491 $\pm$ 0.013 & 0.426 $\pm$ 0.014 & 0.455 $\pm$ 0.012 & 0.452 $\pm$ 0.012 & 0.387 $\pm$ 0.010 & 0.456 $\pm$ 0.013 & 0.455 $\pm$ 0.012 \\
\OIII\ & 4363 & 0.017 $\pm$ 0.003 & 0.051 $\pm$ 0.001 & 0.057 $\pm$ 0.006 & 0.028 $\pm$ 0.002 & 0.018 $\pm$ 0.002 & --- & 0.030 $\pm$ 0.003 & 0.013 $\pm$ 0.002 \\
He I & 4472 & 0.035 $\pm$ 0.003 & 0.042 $\pm$ 0.001 & 0.025 $\pm$ 0.006 & 0.050 $\pm$ 0.002 & 0.050 $\pm$ 0.002 & 0.033 $\pm$ 0.004 & 0.048 $\pm$ 0.003 & 0.045 $\pm$ 0.002 \\
He II & 4687 & --- & 0.006 $\pm$ 0.001 & --- & --- & 0.004 $\pm$ 0.001 & --- & --- & --- \\
\ArIV\ & 4711 & --- & 0.006 $\pm$ 0.001 & --- & --- & --- & --- & --- & --- \\
\ArIV\ & 4740 & --- & --- & --- & --- & --- & --- & --- & --- \\
\HB\ & 4861 & 1.000 $\pm$ 0.022 & 1.000 $\pm$ 0.021 & 1.000 $\pm$ 0.024 & 1.000 $\pm$ 0.022 & 1.000 $\pm$ 0.021 & 1.000 $\pm$ 0.023 & 1.000 $\pm$ 0.023 & 1.000 $\pm$ 0.021 \\
He I & 4922 & --- & 0.009 $\pm$ 0.001 & --- & --- & --- & --- & --- & --- \\
\OIII\ & 4959 & 0.862 $\pm$ 0.019 & 1.357 $\pm$ 0.028 & 0.446 $\pm$ 0.012 & 0.952 $\pm$ 0.021 & 0.701 $\pm$ 0.015 & 0.293 $\pm$ 0.008 & 0.883 $\pm$ 0.020 & 0.592 $\pm$ 0.013 \\
\OIII\ & 5007 & 2.589 $\pm$ 0.057 & 4.091 $\pm$ 0.086 & 1.294 $\pm$ 0.030 & 2.873 $\pm$ 0.062 & 2.091 $\pm$ 0.045 & 0.867 $\pm$ 0.020 & 2.644 $\pm$ 0.059 & 1.774 $\pm$ 0.038 \\
\NI\ & 5199 & 0.006 $\pm$ 0.003 & 0.017 $\pm$ 0.001 & --- & 0.034 $\pm$ 0.002 & --- & --- & 0.028 $\pm$ 0.003 & --- \\
\NII\ & 5755 & --- & 0.011 $\pm$ 0.001 & --- & --- & --- & --- & --- & --- \\
He I & 5876 & 0.107 $\pm$ 0.004 & 0.117 $\pm$ 0.004 & 0.092 $\pm$ 0.005 & 0.119 $\pm$ 0.004 & 0.107 $\pm$ 0.004 & 0.087 $\pm$ 0.004 & 0.112 $\pm$ 0.004 & 0.117 $\pm$ 0.004 \\
\OI\ & 6300 & 0.074 $\pm$ 0.004 & 0.051 $\pm$ 0.002 & 0.090 $\pm$ 0.005 & 0.057 $\pm$ 0.003 & 0.057 $\pm$ 0.002 & 0.086 $\pm$ 0.004 & 0.067 $\pm$ 0.003 & 0.047 $\pm$ 0.002 \\
\SIII\ & 6312 & 0.020 $\pm$ 0.002 & 0.012 $\pm$ 0.001 & --- & 0.010 $\pm$ 0.001 & 0.010 $\pm$ 0.001 & --- & 0.013 $\pm$ 0.002 & 0.009 $\pm$ 0.001 \\
\OI\ & 6364 & 0.024 $\pm$ 0.002 & 0.016 $\pm$ 0.001 & 0.031 $\pm$ 0.003 & 0.023 $\pm$ 0.001 & 0.015 $\pm$ 0.001 & 0.024 $\pm$ 0.003 & 0.014 $\pm$ 0.002 & 0.014 $\pm$ 0.001 \\
\NII\ & 6548 & 0.084 $\pm$ 0.004 & 0.090 $\pm$ 0.004 & 0.385 $\pm$ 0.017 & 0.100 $\pm$ 0.005 & 0.161 $\pm$ 0.007 & 0.319 $\pm$ 0.008 & 0.109 $\pm$ 0.005 & 0.143 $\pm$ 0.006 \\
\HA\ & 6563 & 2.858 $\pm$ 0.123 & 2.821 $\pm$ 0.121 & 2.732 $\pm$ 0.123 & 2.837 $\pm$ 0.124 & 2.845 $\pm$ 0.122 & 2.705 $\pm$ 0.059 & 2.827 $\pm$ 0.123 & 2.854 $\pm$ 0.124 \\
\NII\ & 6583 & 0.259 $\pm$ 0.012 & 0.230 $\pm$ 0.010 & 1.110 $\pm$ 0.051 & 0.316 $\pm$ 0.014 & 0.492 $\pm$ 0.021 & 0.981 $\pm$ 0.022 & 0.328 $\pm$ 0.015 & 0.416 $\pm$ 0.018 \\
He I & 6678 & 0.030 $\pm$ 0.002 & 0.028 $\pm$ 0.001 & 0.015 $\pm$ 0.003 & 0.026 $\pm$ 0.002 & 0.033 $\pm$ 0.002 & 0.018 $\pm$ 0.003 & 0.021 $\pm$ 0.002 & 0.029 $\pm$ 0.002 \\
\SII\ & 6717 & 0.367 $\pm$ 0.017 & 0.207 $\pm$ 0.010 & 0.414 $\pm$ 0.020 & 0.315 $\pm$ 0.015 & 0.348 $\pm$ 0.016 & 0.435 $\pm$ 0.010 & 0.381 $\pm$ 0.018 & 0.315 $\pm$ 0.015 \\
\SII\ & 6731 & 0.255 $\pm$ 0.012 & 0.154 $\pm$ 0.007 & 0.332 $\pm$ 0.016 & 0.231 $\pm$ 0.011 & 0.262 $\pm$ 0.012 & 0.317 $\pm$ 0.008 & 0.272 $\pm$ 0.013 & 0.237 $\pm$ 0.011 \\
He I & 7065 & 0.016 $\pm$ 0.002 & 0.028 $\pm$ 0.001 & --- & 0.017 $\pm$ 0.001 & 0.013 $\pm$ 0.001 & --- & 0.019 $\pm$ 0.002 & 0.019 $\pm$ 0.001 \\
\ArIII\ & 7136 & 0.082 $\pm$ 0.005 & 0.071 $\pm$ 0.004 & --- & 0.056 $\pm$ 0.003 & 0.018 $\pm$ 0.001 & 0.049 $\pm$ 0.003 & 0.079 $\pm$ 0.005 & 0.060 $\pm$ 0.003 \\
\OII\ & 7319 & --- & --- & --- & 0.031 $\pm$ 0.002 & 0.017 $\pm$ 0.001 & --- & 0.035 $\pm$ 0.003 & --- \\
\OII\ B & 7325 & --- & 0.038 $\pm$ 0.002 & --- & 0.045 $\pm$ 0.003 & 0.034 $\pm$ 0.002 & --- & 0.057 $\pm$ 0.004 & 0.040 $\pm$ 0.002 \\
\OII\ & 7330 & --- & --- & --- & 0.008 $\pm$ 0.001 & 0.017 $\pm$ 0.001 & --- & 0.022 $\pm$ 0.002 & --- \\
\SIII\ & 9069 & --- & 0.145 $\pm$ 0.011 & --- & 0.198 $\pm$ 0.016 & 0.168 $\pm$ 0.013 & 0.383 $\pm$ 0.009 & 0.205 $\pm$ 0.016 & 0.244 $\pm$ 0.019 \\
\SIII\ & 9531 & --- & 0.358 $\pm$ 0.029 & --- & 0.424 $\pm$ 0.035 & --- & --- & 0.359 $\pm$ 0.029 & --- \\\enddata
\label{tab:LineRatiosCOR}
\end{deluxetable*}
\end{turnpage}

\begin{turnpage}
\begin{deluxetable*}{ccccccccc}
\tabletypesize{\scriptsize}
\tablenum{2}
\tablewidth{0pt}
\tablecaption{Emission-Line Intensity Ratios Relative to \HB\ for Observed Galaxies (2 of 2)}
\tablehead{\colhead{Ion}&\colhead{$\lambda$ [\AA]}&\colhead{KISSR 1056}&\colhead{KISSR 1379}&\colhead{KISSR 1734}&\colhead{KISSR 1955}&\colhead{KISSR 2117}&\colhead{KISSR 2125}&\colhead{KISSR 2132}}
\startdata
\OII\ & 3727 & 3.204 $\pm$ 0.133 & 3.022 $\pm$ 0.125 & 1.806 $\pm$ 0.073 & 3.437 $\pm$ 0.150 & 3.466 $\pm$ 0.142 & 3.490 $\pm$ 0.156 & 4.188 $\pm$ 0.174 \\
H 10 & 3798 & 0.029 $\pm$ 0.003 & --- & 0.042 $\pm$ 0.002 & --- & 0.048 $\pm$ 0.002 & --- & --- \\
H 9 & 3835 & 0.058 $\pm$ 0.004 & 0.048 $\pm$ 0.003 & 0.065 $\pm$ 0.003 & --- & 0.069 $\pm$ 0.003 & --- & --- \\
\NeIII\ & 3869 & 0.272 $\pm$ 0.011 & 0.188 $\pm$ 0.007 & 0.489 $\pm$ 0.018 & 0.095 $\pm$ 0.008 & 0.168 $\pm$ 0.007 & 0.068 $\pm$ 0.005 & 0.250 $\pm$ 0.010 \\
He I+H 8 & 3889 & 0.156 $\pm$ 0.007 & 0.185 $\pm$ 0.007 & 0.189 $\pm$ 0.007 & 0.178 $\pm$ 0.010 & 0.195 $\pm$ 0.008 & 0.134 $\pm$ 0.007 & 0.168 $\pm$ 0.008 \\
\NeIII+\HE\ & 3970 & 0.182 $\pm$ 0.007 & 0.162 $\pm$ 0.006 & 0.303 $\pm$ 0.010 & 0.113 $\pm$ 0.008 & 0.208 $\pm$ 0.007 & 0.099 $\pm$ 0.005 & 0.151 $\pm$ 0.007 \\
He I & 4026 & --- & --- & 0.016 $\pm$ 0.001 & --- & 0.015 $\pm$ 0.001 & --- & --- \\
\SII\ B & 4073 & 0.037 $\pm$ 0.003 & 0.019 $\pm$ 0.002 & 0.025 $\pm$ 0.001 & 0.049 $\pm$ 0.006 & 0.021 $\pm$ 0.001 & 0.029 $\pm$ 0.003 & 0.063 $\pm$ 0.004 \\
\HD\ & 4102 & 0.238 $\pm$ 0.008 & 0.256 $\pm$ 0.008 & 0.259 $\pm$ 0.008 & 0.235 $\pm$ 0.011 & 0.278 $\pm$ 0.009 & 0.226 $\pm$ 0.008 & 0.232 $\pm$ 0.008 \\
\HG\ & 4340 & 0.452 $\pm$ 0.013 & 0.478 $\pm$ 0.013 & 0.473 $\pm$ 0.012 & 0.439 $\pm$ 0.015 & 0.492 $\pm$ 0.013 & 0.461 $\pm$ 0.013 & 0.450 $\pm$ 0.013 \\
\OIII\ & 4363 & 0.027 $\pm$ 0.002 & 0.019 $\pm$ 0.002 & 0.076 $\pm$ 0.002 & --- & 0.015 $\pm$ 0.001 & 0.020 $\pm$ 0.003 & 0.028 $\pm$ 0.003 \\
He I & 4472 & 0.050 $\pm$ 0.003 & 0.055 $\pm$ 0.002 & 0.042 $\pm$ 0.001 & 0.077 $\pm$ 0.006 & 0.039 $\pm$ 0.002 & 0.043 $\pm$ 0.003 & 0.029 $\pm$ 0.003 \\
He II & 4687 & --- & 0.009 $\pm$ 0.001 & 0.009 $\pm$ 0.001 & --- & --- & --- & --- \\
\ArIV\ & 4711 & --- & 0.004 $\pm$ 0.001 & 0.009 $\pm$ 0.001 & 0.009 $\pm$ 0.005 & --- & --- & --- \\
\ArIV\ & 4740 & --- & --- & 0.004 $\pm$ 0.001 & --- & --- & --- & --- \\
\HB\ & 4861 & 1.000 $\pm$ 0.022 & 1.000 $\pm$ 0.021 & 1.000 $\pm$ 0.021 & 1.000 $\pm$ 0.025 & 1.000 $\pm$ 0.022 & 1.000 $\pm$ 0.021 & 1.000 $\pm$ 0.022 \\
He I & 4922 & --- & --- & 0.008 $\pm$ 0.001 & --- & 0.009 $\pm$ 0.001 & --- & --- \\
\OIII\ & 4959 & 1.027 $\pm$ 0.023 & 0.803 $\pm$ 0.017 & 1.666 $\pm$ 0.035 & 0.258 $\pm$ 0.008 & 0.859 $\pm$ 0.019 & 0.253 $\pm$ 0.006 & 0.925 $\pm$ 0.021 \\
\OIII\ & 5007 & 3.051 $\pm$ 0.067 & 2.389 $\pm$ 0.051 & 4.969 $\pm$ 0.105 & 0.795 $\pm$ 0.020 & 2.593 $\pm$ 0.056 & 0.744 $\pm$ 0.016 & 2.776 $\pm$ 0.061 \\
\NI\ & 5199 & --- & 0.023 $\pm$ 0.001 & --- & 0.022 $\pm$ 0.005 & 0.021 $\pm$ 0.001 & 0.044 $\pm$ 0.002 & 0.026 $\pm$ 0.003 \\
\NII\ & 5755 & --- & --- & 0.007 $\pm$ 0.001 & --- & 0.006 $\pm$ 0.001 & --- & --- \\
He I & 5876 & 0.108 $\pm$ 0.004 & 0.119 $\pm$ 0.004 & 0.117 $\pm$ 0.004 & 0.109 $\pm$ 0.005 & 0.103 $\pm$ 0.004 & 0.098 $\pm$ 0.004 & 0.114 $\pm$ 0.005 \\
\OI\ & 6300 & 0.078 $\pm$ 0.003 & 0.043 $\pm$ 0.002 & 0.048 $\pm$ 0.002 & 0.061 $\pm$ 0.004 & 0.053 $\pm$ 0.002 & 0.054 $\pm$ 0.002 & 0.064 $\pm$ 0.003 \\
\SIII\ & 6312 & --- & 0.011 $\pm$ 0.001 & 0.015 $\pm$ 0.001 & 0.011 $\pm$ 0.003 & 0.009 $\pm$ 0.001 & 0.005 $\pm$ 0.001 & 0.008 $\pm$ 0.002 \\
\OI\ & 6364 & 0.028 $\pm$ 0.002 & 0.011 $\pm$ 0.001 & 0.015 $\pm$ 0.001 & 0.016 $\pm$ 0.003 & 0.016 $\pm$ 0.001 & 0.019 $\pm$ 0.001 & 0.015 $\pm$ 0.002 \\
\NII\ & 6548 & 0.103 $\pm$ 0.005 & 0.103 $\pm$ 0.005 & 0.097 $\pm$ 0.004 & 0.222 $\pm$ 0.011 & 0.141 $\pm$ 0.006 & 0.225 $\pm$ 0.010 & 0.095 $\pm$ 0.005 \\
\HA\ & 6563 & 2.843 $\pm$ 0.123 & 2.849 $\pm$ 0.123 & 2.808 $\pm$ 0.119 & 2.860 $\pm$ 0.129 & 2.867 $\pm$ 0.123 & 2.772 $\pm$ 0.130 & 2.835 $\pm$ 0.123 \\
\NII\ & 6583 & 0.325 $\pm$ 0.014 & 0.311 $\pm$ 0.014 & 0.277 $\pm$ 0.012 & 0.680 $\pm$ 0.031 & 0.423 $\pm$ 0.018 & 0.708 $\pm$ 0.034 & 0.305 $\pm$ 0.014 \\
He I & 6678 & 0.024 $\pm$ 0.002 & 0.031 $\pm$ 0.002 & 0.032 $\pm$ 0.001 & 0.026 $\pm$ 0.003 & 0.024 $\pm$ 0.001 & 0.020 $\pm$ 0.001 & 0.034 $\pm$ 0.003 \\
\SII\ & 6717 & 0.361 $\pm$ 0.017 & 0.286 $\pm$ 0.013 & 0.176 $\pm$ 0.008 & 0.572 $\pm$ 0.028 & 0.299 $\pm$ 0.014 & 0.424 $\pm$ 0.022 & 0.400 $\pm$ 0.019 \\
\SII\ & 6731 & 0.268 $\pm$ 0.013 & 0.207 $\pm$ 0.010 & 0.136 $\pm$ 0.006 & 0.414 $\pm$ 0.021 & 0.226 $\pm$ 0.011 & 0.305 $\pm$ 0.016 & 0.289 $\pm$ 0.014 \\
He I & 7065 & --- & 0.017 $\pm$ 0.001 & 0.038 $\pm$ 0.002 & 0.010 $\pm$ 0.003 & 0.033 $\pm$ 0.002 & 0.009 $\pm$ 0.001 & --- \\
\ArIII\ & 7136 & --- & 0.073 $\pm$ 0.004 & 0.053 $\pm$ 0.003 & 0.070 $\pm$ 0.005 & 0.081 $\pm$ 0.004 & 0.032 $\pm$ 0.002 & --- \\
\OII\ & 7319 & --- & 0.022 $\pm$ 0.001 & 0.022 $\pm$ 0.001 & --- & --- & 0.017 $\pm$ 0.001 & --- \\
\OII\ B & 7325 & --- & 0.042 $\pm$ 0.002 & 0.041 $\pm$ 0.002 & --- & 0.069 $\pm$ 0.004 & 0.026 $\pm$ 0.002 & --- \\
\OII\ & 7330 & --- & 0.020 $\pm$ 0.001 & 0.018 $\pm$ 0.001 & --- & --- & 0.013 $\pm$ 0.001 & --- \\
\SIII\ & 9069 & --- & 0.125 $\pm$ 0.010 & 0.122 $\pm$ 0.009 & --- & 0.288 $\pm$ 0.022 & 0.133 $\pm$ 0.012 & --- \\
\SIII\ & 9531 & --- & 0.238 $\pm$ 0.019 & --- & --- & 0.385 $\pm$ 0.031 & 0.281 $\pm$ 0.025 & --- \\\enddata
\label{tab:LineRatiosCOR}
\end{deluxetable*}
\end{turnpage}
\clearpage

\begin{deluxetable*}{ccccccc}
\tabletypesize{\tiny}
\tablewidth{0pt}
\tablenum{3}
\tablecaption{Electron Densities, Temperatures, and Ionic Abundances (1 of 2)}
\tablehead{\colhead{KISSR}&\colhead{\OIII\ $t_{e}$ [K]}&\colhead{\SII\ $n_{e}$ [cm$^{-3}$]}&\colhead{log(He/H)+12}&\colhead{log(N/H)+12}&\colhead{log(N/O)}&\colhead{log(O/H)+12}}
\startdata
0148 & 10120 $\pm$ 343 & 100 $\pm$ --- & 10.91 $\pm$ 0.05 & 6.94 $\pm$ 0.05 & -1.30 $\pm$ 0.08 & 8.23 $\pm$ 0.06 \\
0242 & 12490 $\pm$ 136 & 68.5 $\pm$ 25.4 & 10.96 $\pm$ 0.01 & 6.99 $\pm$ 0.02 & -0.95 $\pm$ 0.02 & 7.93 $\pm$ 0.02 \\
0258 & 23730 $\pm$ 2894 & 176 $\pm$ 38.1 & 10.98 $\pm$ 0.12 & 6.97 $\pm$ 0.04 & -0.42 $\pm$ 0.02 & 7.38 $\pm$ 0.03 \\
0451 & 11390 $\pm$ 403 & 48.8 $\pm$ 24.7 & 10.96 $\pm$ 0.01 & 6.91 $\pm$ 0.04 & -1.20 $\pm$ 0.02 & 8.12 $\pm$ 0.04 \\
0475 & 10920 $\pm$ 500 & 84.8 $\pm$ 28.3 & 10.91 $\pm$ 0.01 & 7.09 $\pm$ 0.04 & -1.00 $\pm$ 0.02 & 8.10 $\pm$ 0.06 \\
0512 & 10000 $\pm$ --- & 49.6 $\pm$ 43.5 & 10.82 $\pm$ 0.04 & 7.43 $\pm$ 0.01 & -0.46 $\pm$ 0.01 & 7.89 $\pm$ 0.01 \\
0590 & 12020 $\pm$ 449 & 15.7 $\pm$ 28.3 & 10.94 $\pm$ 0.04 & 6.89 $\pm$ 0.04 & -1.13 $\pm$ 0.02 & 8.02 $\pm$ 0.05 \\
0653 & 10380 $\pm$ 335 & 84.1 $\pm$ 28.7 & 10.95 $\pm$ 0.01 & 7.08 $\pm$ 0.03 & -1.00 $\pm$ 0.02 & 8.08 $\pm$ 0.04 \\
1056 & 11050 $\pm$ 380 & 66.0 $\pm$ 23.6 & --- & 7.00 $\pm$ 0.04 & -1.14 $\pm$ 0.02 & 8.14 $\pm$ 0.04 \\
1379 & 10730 $\pm$ 516 & 39.5 $\pm$ 28.1 & 10.96 $\pm$ 0.01 & 6.98 $\pm$ 0.05 & -1.13 $\pm$ 0.02 & 8.11 $\pm$ 0.06 \\
1734 & 13610 $\pm$ 154 & 114 $\pm$ 34.0 & 10.96 $\pm$ 0.01 & 7.04 $\pm$ 0.03 & -0.91 $\pm$ 0.03 & 7.95 $\pm$ 0.02 \\
1955 & 10000 $\pm$ --- & 41.4 $\pm$ 31.4 & 10.92 $\pm$ 0.04 & 7.18 $\pm$ 0.01 & -0.89 $\pm$ 0.02 & 8.06 $\pm$ 0.01 \\
2117 & 9751 $\pm$ 295 & 91.0 $\pm$ 26.5 & 10.89 $\pm$ 0.04 & 7.18 $\pm$ 0.03 & -1.15 $\pm$ 0.03 & 8.33 $\pm$ 0.04 \\
2125 & 10000 $\pm$ --- & 100 $\pm$ --- & 10.91 $\pm$ 0.04 & 6.67 $\pm$ 0.02 & -0.67 $\pm$ 0.02 & 7.34 $\pm$ 0.03 \\
2132 & 11590 $\pm$ 417 & 31.5 $\pm$ 26.1 & 10.95 $\pm$ 0.02 & 6.85 $\pm$ 0.04 & -1.27 $\pm$ 0.02 & 8.12 $\pm$ 0.04 \\
\enddata
\label{tab:ionic_abundances}
\end{deluxetable*}

\begin{deluxetable*}{ccccccc}
\tabletypesize{\tiny}
\tablewidth{0pt}
\tablenum{3}
\tablecaption{Electron Densities, Temperatures, and Ionic Abundances (2 of 2)}
\tablehead{\colhead{KISSR}&\colhead{12+log(Ne/H)}&\colhead{log(Ne/O)}&\colhead{12+log(S/H)}&\colhead{log(S/O)}&\colhead{12+log(Ar/H)}&\colhead{log(Ar/O)}}
\startdata
0148 & 7.57 $\pm$ 0.09 & -0.67 $\pm$ 0.04 & 6.81 $\pm$ 0.07 & -1.42 $\pm$ 0.07 & 5.88 $\pm$ 0.06 & -2.36 $\pm$ 0.08 \\
0242 & 7.33 $\pm$ 0.02q & -0.61 $\pm$ 0.01 & 6.35 $\pm$ 0.02 & -1.59 $\pm$ 0.02 & 5.69 $\pm$ 0.02 & -2.25 $\pm$ 0.01 \\
0258 & 6.68 $\pm$ 0.05 & -0.70 $\pm$ 0.04 & 5.75 $\pm$ 0.05 & -1.63 $\pm$ 0.02 & --- & --- \\
0451 & 7.45 $\pm$ 0.05 & -0.67 $\pm$ 0.01 & 6.60 $\pm$ 0.03 & -1.52 $\pm$ 0.04 & 5.59 $\pm$ 0.04 & -2.53 $\pm$ 0.03 \\
0475 & 7.43 $\pm$ 0.07 & -0.67 $\pm$ 0.01 & 6.54 $\pm$ 0.04 & -1.55 $\pm$ 0.04 & 5.14 $\pm$ 0.06 & -2.95 $\pm$ 0.04 \\
0512 & 7.11 $\pm$ 0.04 & -0.78 $\pm$ 0.04 & 6.85 $\pm$ 0.01 & -1.04 $\pm$ 0.01 & 5.67 $\pm$ 0.04 & -2.23 $\pm$ 0.04 \\
0590 & 7.33 $\pm$ 0.06 & -0.68 $\pm$ 0.02 & 6.55 $\pm$ 0.03 & -1.47 $\pm$ 0.04 & 5.69 $\pm$ 0.06 & -2.33 $\pm$ 0.05 \\
0653 & 7.28 $\pm$ 0.05 & -0.80 $\pm$ 0.02 & 6.77 $\pm$ 0.03 & -1.31 $\pm$ 0.04 & 5.71 $\pm$ 0.04 & -2.37 $\pm$ 0.02 \\
1056 & 7.51 $\pm$ 0.05 & -0.63 $\pm$ 0.02 & 6.05 $\pm$ 0.03 & -2.09 $\pm$ 0.02 & --- & --- \\
1379 & 7.43 $\pm$ 0.07 & -0.68 $\pm$ 0.02 & 6.38 $\pm$ 0.04 & -1.73 $\pm$ 0.04 & 5.76 $\pm$ 0.05 & -2.35 $\pm$ 0.02 \\
1734 & 7.32 $\pm$ 0.02 & -0.63 $\pm$ 0.01 & 6.17 $\pm$ 0.02 & -1.78 $\pm$ 0.02 & 5.40 $\pm$ 0.02 & -2.55 $\pm$ 0.02 \\
1955 & 7.59 $\pm$ 0.05 & -0.48 $\pm$ 0.04 & 6.70 $\pm$ 0.03 & -1.37 $\pm$ 0.03 & 5.81 $\pm$ 0.04 & -2.25 $\pm$ 0.05 \\
2117 & 7.59 $\pm$ 0.05 & -0.73 $\pm$ 0.02 & 6.87 $\pm$ 0.03 & -1.46 $\pm$ 0.04 & 5.91 $\pm$ 0.06 & -2.42 $\pm$ 0.06 \\
2125 & 6.64 $\pm$ 0.05 & -0.70 $\pm$ 0.03 & 6.06 $\pm$ 0.02 & -1.28 $\pm$ 0.02 & 4.91 $\pm$ 0.03 & -2.43 $\pm$ 0.03 \\
2132 & 7.48 $\pm$ 0.05 & -0.64 $\pm$ 0.02 & 6.05 $\pm$ 0.03 & -2.08 $\pm$ 0.02 & --- & --- \\
\enddata
\label{tab:ionic_abundances}
\end{deluxetable*}

\subsubsection{Sulfur \Te} 

\begin{figure*}
\plotone{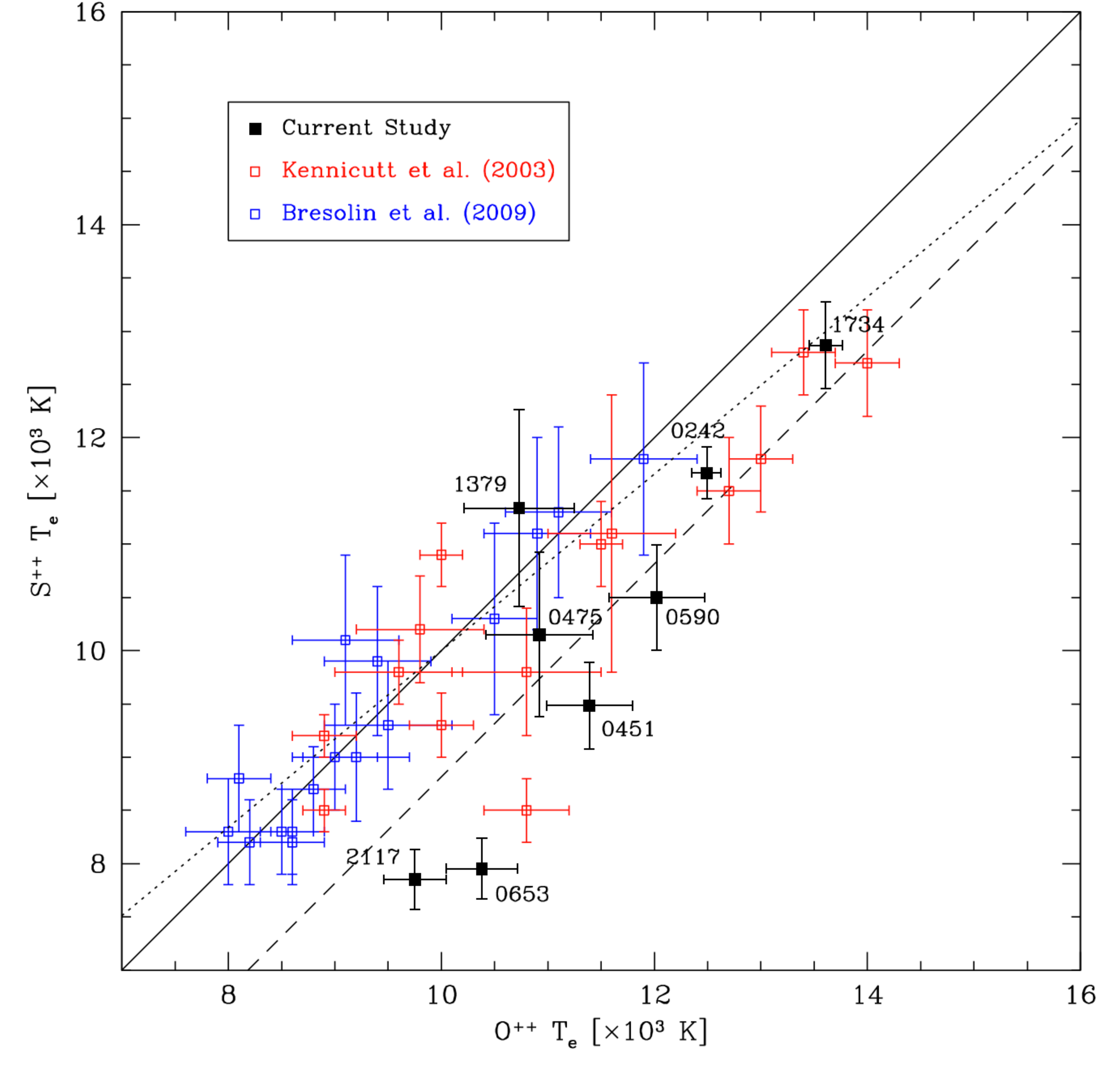}
\vspace{-0.5cm}
\caption{Comparison of \Te\ calculated using O$^{++}$ and S$^{++}$ for galaxies where all necessary emission lines are measured.
The solid line marks the one-to-one equivalence, while the dashed line represents an empirically derived offset of 1184 K to lower \Te\ based upon the average difference between the measured O$^{++}$ and S$^{++}$ electron temperatures.
The dotted line marks the model prediction fit from \citet{bib:Garnett1992}.
The red and blue points represent \HII\ regions in M101 and NGC 300 from \citet{bib:Kennicutt2003} and \citet{bib:Bresolin2009}, respectively, over a similar O$^{++}$ \Te\ range.}
\label{fig:OvsS_final}
\vspace{0.5cm}
\end{figure*}

\indent Determination of S$^{++}$ temperatures requires observation of the near-infrared \SIII$\lambda\lambda$9069,9531 doublet and the temperature-sensitive auroral \SIII$\lambda$6312 line \citep{bib:Garnett1989}.
These NIR sulfur emission lines become increasingly important to nebular cooling at lower temperatures \citep{bib:DiazPerez-Montero2000}.
While they benefit from their being bright and relatively free from reddening effects, NIR sulfur requires observations that reach to substantially longer wavelengths than are commonly used for nebular abundance work.
The $\lambda$6312 line is generally weaker than $\lambda$4363, and can suffer blending due to proximity to the \OI\ night-sky line at 6300 \AA\ as well as nebular \OI\ emission, however spectrograph CCDs are often more efficient at redder wavelengths.
In addition, $\lambda$4363 can be subject to significant reddening effects by the presence of dust that impedes measurement, particularly for metal-rich nebulae in which the line strength is naturally diminished.
A \Te\ measurement utilizing \SIII$\lambda$6312 is comparatively less influenced by dust and could remain viable for a wider range of physical conditions.
Additionally, while the strength of both the $\lambda$4363 and $\lambda$6312 auroral lines weakens significantly at high metallicity, the studies of \citet{bib:Bresolin2004} and \citet{bib:Bresolin2005} have shown that the $\lambda$6312 line remains a viable alternative for the determination of nebular \Te\ to somewhat higher abundances than $\lambda$4363.\\
\indent For ten of our fifteen galaxies, the red spectral grating was switched to observe the nebular NIR \SIII\ doublet, both of which were detected in six objects.
In the other four cases, the $\lambda$9531 line was redshifted off of the CCD, and its flux value was estimated from the strength of the $\lambda$9069 line and the theoretical ratio of $\lambda$9531/$\lambda$9069 = 2.44 \citep{bib:MendozaZeippen1982}.
The weak auroral $\lambda$6312 line was measured in twelve of our fifteen galaxies, including nine of which possessed NIR spectra.
KISSR 0512 alone possessed NIR \SIII\ data without recovery of the $\lambda$6312 auroral line.
The \OIII$\lambda$4363 line was not measured for KISSR 2125, and so a comparison of O$^{++}$ and S$^{++}$ \Te-method abundances was not possible in this case.
This reduces the total count of galaxies for which NIR \SIII, \SIII$\lambda$6312, and \OIII$\lambda$4363 were all measured to eight:
KISSR 0242, KISSR 0451, KISSR 0475, KISSR 0590, KISSR 0653, KISSR 1379, KISSR 1734, and KISSR 2117.\\
\indent The \texttt{ELSA} program was used to compute a S$^{++}$ temperature for the objects that possessed the necessary lines.
Figure \ref{fig:OvsS_final} compares the \Te\ calculated using O$^{++}$ and S$^{++}$ lines.
The solid line indicates the one-to-one equivalence, while the dashed line represents an empirically derived offset of 1184 K to lower \Te\ based upon the average difference between the measured O$^{++}$ and S$^{++}$ electron temperatures.
The dotted line represents the fit from \citet{bib:Garnett1992}, which uses photoionization models to relate electron temperatures of the different ionization zones, with functional form:
\[
t_{e}(\text{S}^{++}) = 0.83\times t_{e}(\text{O}^{++}) + 1700\ {\text K}.
\]
We find that in all but one case, the measured electron temperature calculated using the \SIII\ lines is lower than the value determined from the \OIII\ lines.\\
\indent Included in Figure \ref{fig:OvsS_final} are red points representing \HII\ region data from the \citet{bib:Kennicutt2003} study of M101 and blue points representing the \HII\ region data from the \citet{bib:Bresolin2009} study of NGC 300.
These data cover a similar range in O$^{++}$ \Te\ as our work.
\citet{bib:Bresolin2009} data points strongly correlate to the equality line, especially at \Te\ below $\sim$10,000 K, with an average offset of 75 K to \emph{higher} electron temperatures using S$^{++}$.
Data from the \citet{bib:Kennicutt2003} study, however, shows fairly good agreement with our offset above $\sim$12,000 K, with an overall average offset of 564 K to lower electron temperatures using S$^{++}$.
The work of \citet{bib:Berg2015} revisits the study of \HII\ regions in M101 by \citet{bib:Kennicutt2003}, finding that an adoption  of updated atomic data substantially decreases the measured \Te\ as determined with S$^{++}$ (by roughly 10\%, or $\sim$1000 K), while the O$^{++}$ remains mostly unchanged (see their Figure 3).
This result appears to agree with the results from our study, although additional investigation is required.

\subsection{Direct-Method Metal Abundances} 

\begin{figure*}
\epsscale{0.736}
\plotone{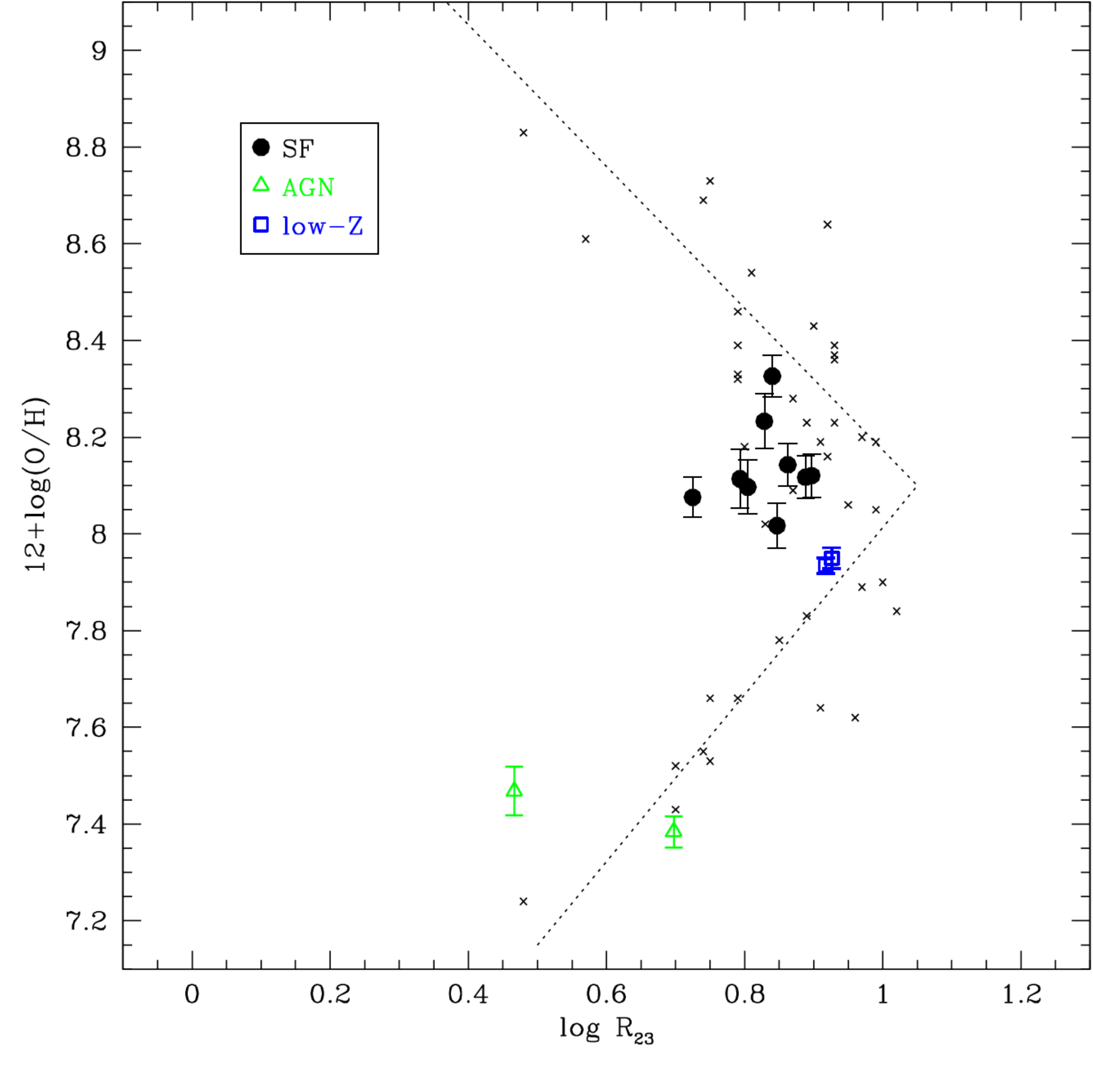}
\vspace{-0.3cm}
\caption{\R\ diagram with direct-method oxygen abundances from this study.
The dotted line represents the \R\ calibration from \citet{bib:EdmundsPagel1984}, while crosses are reference points from \citet{bib:McGaugh1991}.
Filled black circles represent the nine regular star-forming galaxies with measured \Te.
Open green triangles represent possible AGN contaminated targets, while open blue squares are low-metallicity, high-luminosity compact star-forming galaxies.  
Both of these anomalous target types are discussed in more detail in \S5.2.}
\label{fig:R23sf_all}
\vspace{-0.3cm}
\end{figure*}

\indent We use the program \texttt{ELSA} \citep{bib:Johnson2006} to calculate ionic abundances relative to hydrogen.
The input data include the densities and temperatures of each ionization zone, as calculated above, and the emission-line ratios previously presented as Table 3.
We calculate the abundance of the following ions with respect to H$^{+}$: O$^{+}$, O$^{++}$, N$^{+}$, S$^{+}$, S$^{++}$, Ne$^{++}$, and Ar$^{++}$.\\
\indent The total oxygen abundance is assumed to be given by
\[
\frac{\text{O}}{\text{H}} = \frac{\text{O}^{+}}{\text{H}^{+}} + \frac{\text{O}^{++}}{\text{H}^{+}},
\]
following the standard practice.
Additional ionization states for other elements that are present in the nebula but do not emit in the optical spectrum are accounted for with ionization correction factors (ICFs).
We use the prescriptions given by \citet{bib:Izotov1994} for the ICFs for N, Ne, S, and Ar:
\[
\text{ICF(N)} = \frac{\text{N}}{\text{N}^{+}} = \frac{\text{O}}{\text{O}^{+}},
\]
\[
\text{ICF(Ne)} = \frac{\text{Ne}}{\text{Ne}^{++}} = \frac{\text{O}}{\text{O}^{++}},
\]
\[
\text{ICF(S)} = \frac{\text{S}}{\text{S}^{+} + \text{S}^{++}} =
\]
\[
\text{0.013 + $x$\{5.10 + $x$[-12.78 + $x$(14.77 - 6.11$x$)]\})$^{-1}$},
\]
\[
\text{ICF(Ar}) = \frac{\text{Ar}}{\text{Ar}^{++}} = [0.15 + x(2.39 - 2.64x)]^{-1},
\]
\[
\text{where}\ x = \frac{\text{O}}{\text{O}^{+}}.
\]
\citet{bib:Izotov1994} suggest that the flux for O$^{+++}$ can be inferred by
\[
\frac{\text{O}^{+++}}{\text{O}^{++}} = \frac{\text{He}^{++}}{\text{He}^{+}}.
\]
In the few cases in which we observe He$^{++}$ in the spectra, however, the line is very noisy, and we assume that the amount of O$^{+++}$ in the nebulae is negligible.\\
\indent Using the ionic abundances and ionization correction factors given above, we calculate the 
He/H, N/H, N/O, O/H, Ne/H, Ne/O, S/H, S/O, Ar/H, and Ar/O
ratios for each galaxy.
Our results are presented in Table 3.
Direct-method oxygen abundances for our galaxies are plotted against \R\ ratios in Figure \ref{fig:R23sf_all}.
Eleven of the fifteen observed galaxies appear to be regular star-forming galaxies with a range of oxygen abundances between 8.02 $\le$ \abun\ $\le$ 8.33, inhabiting the turnaround region of \R.
The remaining four display incongruous characteristics and will be discussed in more detail in \S5.2.
For two of the regular star-forming galaxies (KISSR 1955 and KISSR 2125), the \OIII$\lambda$4363 line was not clearly detected.
Hence we are unable to provide oxygen-derived direct-method abundances for these two targets.\\
\indent For all eight galaxies with an estimate of \Te\ derived using \SIII\ emission lines, a S$^{++}$ direct-method oxygen abundance was calculated using \texttt{ELSA}.
As expected, the metallicities determined from sulfur-derived \Te\ are higher than those determined from oxygen-derived \Te\ where the S$^{++}$ temperature is lower than that of O$^{++}$ (seven of eight cases).
We return to the issue of metallicities from \SIII\ emission lines in \S5.1.2.

\subsection{Strong Emission-Line (SEL) Method Abundances} 

\begin{figure*}
\epsscale{0.736}
\plotone{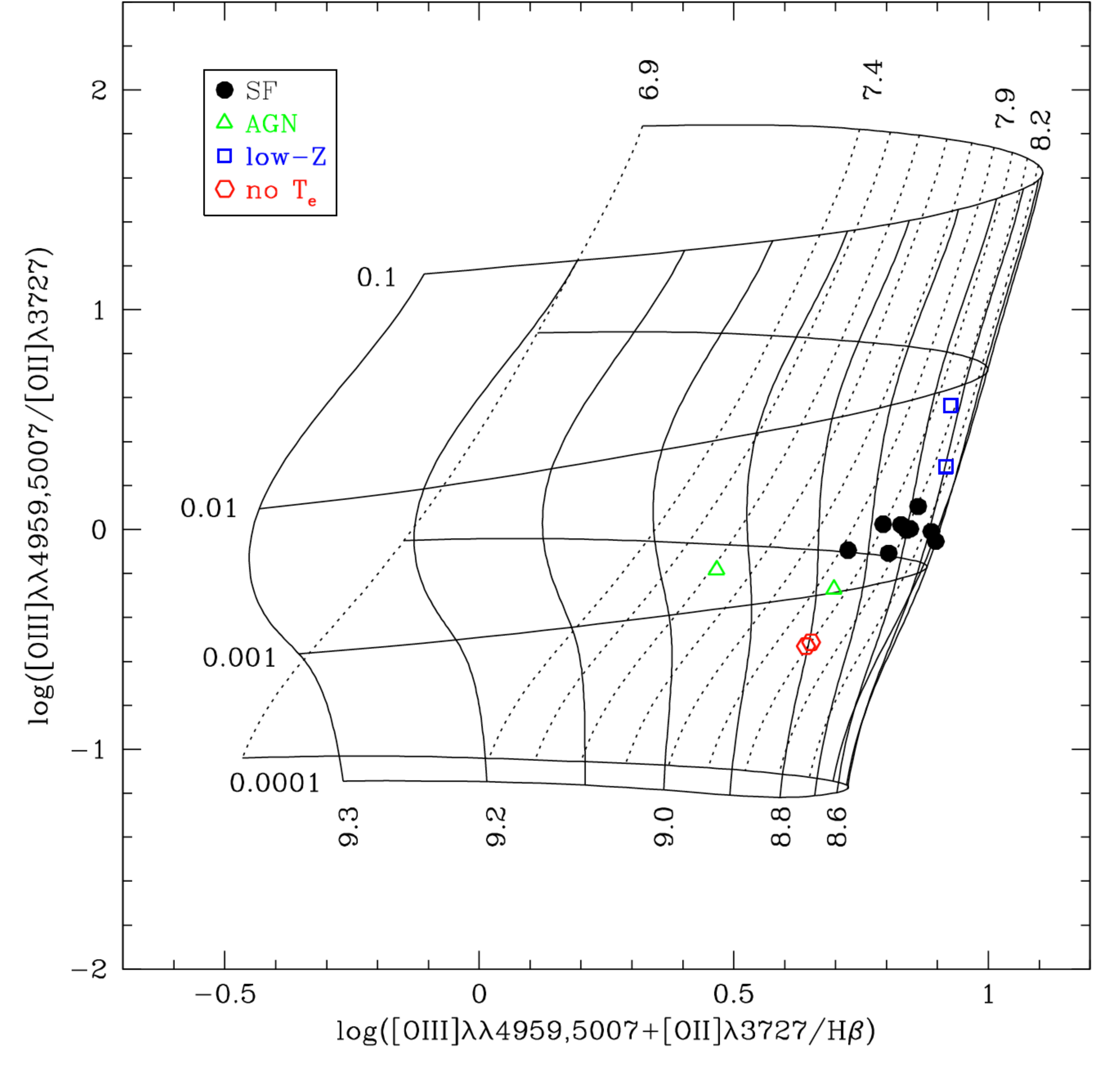}
\vspace{-0.3cm}
\caption{McGaugh model abundance grid for a 60 $M_{\odot}$ IMF upper mass cutoff \citep{bib:McGaugh1991} including the fifteen galaxies from our sample.
The filled black circles represent the nine regular star-forming galaxies for which \OIII$\lambda$4363 was measured.
Green open triangles have potential AGN spectral contamination.
Blue open squares are low-metallicity, high-luminosity compact objects.
Red open hexagons are regular star-forming galaxies that lack reliable \Te\ measurement.
These will be discussed further in \S5.2.
While the \R\ ratio on the x-axis is sensitive to abundance (marked by vertical lines on the grid ranging from 6.9 to 9.3), the $O_{23}$ ratio on the y-axis is sensitive to levels of excitation (marked by horizontal lines on the grid ranging from 0.0001 to 0.1).
The black grid represents photoionization models calculated from \texttt{CLOUDY} \citep{bib:FerlandTruran1981}.
Errors in the line ratio measurements are comparable in size to the plot symbols.}
\label{fig:McGaugh_all}
\vspace{0.5cm}
\end{figure*}

\indent The \citet{bib:McGaugh1991} model grid was derived using photoionization model calculations from \texttt{CLOUDY} \citep{bib:FerlandTruran1981} and employs the abundance-sensitive \R\ line ratio alongside the excitation-sensitive $O_{23}$ line ratio, defined as \OIII$\lambda\lambda$4959,5007/\OII$\lambda$3727 and incorporates an upper- and lower-metallicity branch along the axis defined by the \R\ line ratio.
Data points representing each of the fifteen galaxies from this sample are shown on Figure \ref{fig:McGaugh_all}, differentiated by object type (to be discussed in \S5.2.).
Typical abundance uncertainties of the McGaugh model grid method are $\sigma$ $\approx$ 0.05 for the low-metallicity branch and $\sigma$ $\approx$ 0.10 for the high-metallicity branch, dominated by the uncertainties in the model stellar radiation fields \citep{bib:McGaugh1991}.
The nine regular star-forming galaxies with measured \Te\ from this study fall between the upper- and lower-metallicity branches of the grid.
Abundances for these systems were calculated to be in the range of 8.07 $\le$ \abun\ $\le$ 8.74, and are presented in Table 4.
The two low-metallicity, high-luminosity compact star-forming galaxies appear to reside within the turnaround region, albeit with noticeably higher excitation.
Conversely, the two galaxies for which $\lambda$4363 emission was undetected exhibit comparatively low-excitation.\\
\indent The coarse abundance method \citep{bib:MelbourneSalzer2002, bib:Salzer2005b} incorporates the \OIII$\lambda$5007/\HB\ and \NII$\lambda$6583/\HA\ line ratios calibrated to a combination of \Te\ abundances and previously published \R-like relations.
These two line ratios are observed in the optical spectrum of most local ELGs, typically have good signal-to-noise ratios, and due to the relative proximity of the individual lines, are fairly insensitive to reddening corrections.
A spectral activity diagnostic diagram shows that in the low-metallicity regime, abundance increases with nitrogen line strength as a smooth, single-valued function.
Conversely, for the high-metallicity regime, abundance is anti-correlated with increasing oxygen line strength.
Typical uncertainties of the coarse abundance method are $\sigma$ $\approx$ 0.15-0.20 \citep{bib:MelbourneSalzer2002, bib:Salzer2005b}.
While not exceptionally precise, coarse abundances can be quite useful for large-scale statistical work.
For the nine regular star-forming galaxies observed for this study, the computed coarse-method metallicities range from 8.29 $\le$ \abun\ $\le$ 8.44, and are listed in Table 4.

\begin{deluxetable*}{cccccccc}
\tabletypesize{\footnotesize}
\tablewidth{0pt}
\tablenum{4}
\tablecaption{Electron Temperature, Oxygen Abundance $Z$ Comparison, and Spectral Activity Type}
\tablehead{\colhead{KISSR}&\colhead{O$^{++}$ \Te\ [K]}&\colhead{S$^{++}$ \Te\ [K]}&\colhead{O$^{++}$ $Z$}&\colhead{S$^{++}$ $Z$}&\colhead{McGaugh $Z$}&\colhead{Coarse $Z$}&\colhead{Type}}
\startdata
0148 & 10120 $\pm$ 343 & --- & 8.23 $\pm$ 0.06 & --- & 8.07 $\pm$ 0.10 & 8.29 $\pm$ 0.20 & SF \\
0242 & 12490 $\pm$ 136 & 11670 $\pm$ 242 & 7.93 $\pm$ 0.02 & 8.02 $\pm$ 0.02 & 8.16 $\pm$ 0.10 & 8.21 $\pm$ 0.20 & low-$Z$ \\
0258 & 23730 $\pm$ 2894 & --- & 7.38 $\pm$ 0.03 & --- & 8.76 $\pm$ 0.10 & 8.67 $\pm$ 0.20 & AGN \\
0451 & 11390 $\pm$ 403 & 9485 $\pm$ 410 & 8.12 $\pm$ 0.04 & 8.35 $\pm$ 0.04 & 8.26 $\pm$ 0.10 & 8.31 $\pm$ 0.20 & SF \\
0475 & 10920 $\pm$ 500 & 10150 $\pm$ 773 & 8.10 $\pm$ 0.06 & 8.18 $\pm$ 0.07 & 8.64 $\pm$ 0.10 & 8.44 $\pm$ 0.20 & SF \\
0512 & 14370 $\pm$ 627 & --- & 7.47 $\pm$ 0.05 & --- & 8.94 $\pm$ 0.10 & 8.71 $\pm$ 0.20 & AGN \\
0590 & 12020 $\pm$ 449 & 10500 $\pm$ 498 & 8.02 $\pm$ 0.05 & 8.18 $\pm$ 0.04 & 8.12 $\pm$ 0.10 & 8.33 $\pm$ 0.20 & SF \\
0653 & 10380 $\pm$ 335 & 7952 $\pm$ 285 & 8.08 $\pm$ 0.04 & 8.43 $\pm$ 0.04 & 8.74 $\pm$ 0.10 & 8.43 $\pm$ 0.20 & SF \\
1056 & 11050 $\pm$ 335 & --- & 8.14 $\pm$ 0.04 & --- & 8.34 $\pm$ 0.10 & 8.31 $\pm$ 0.20 & SF \\
1379 & 10730 $\pm$ 516 & 11340 $\pm$ 920 & 8.11 $\pm$ 0.06 & 8.05 $\pm$ 0.05 & 8.33 $\pm$ 0.10 & 8.34 $\pm$ 0.20 & SF \\
1734 & 13610 $\pm$ 154 & 12870 $\pm$ 406 & 7.95 $\pm$ 0.02 & 8.01 $\pm$ 0.02 & 8.07 $\pm$ 0.10 & 8.21 $\pm$ 0.20 & low-$Z$ \\
1955 & --- & --- & --- & --- & 8.79 $\pm$ 0.10 & 8.64 $\pm$ 0.20 & no \Te\ \\
2117 & 9751 $\pm$ 295 & 7851 $\pm$ 278 & 8.33 $\pm$ 0.04 & 8.55 $\pm$ 0.04 & 8.36 $\pm$ 0.10 & 8.38 $\pm$ 0.20 & SF \\
2125 & --- & 7819 $\pm$ 924 & --- & --- & 8.79 $\pm$ 0.10 & 8.66 $\pm$ 0.20 & no \Te\ \\
2132 & 11590 $\pm$ 417 & --- & 8.12 $\pm$ 0.04 & --- & 8.35 $\pm$ 0.10 & 8.31 $\pm$ 0.20 & SF \\
\enddata
\label{tab:abun_comp}
\end{deluxetable*}

\section{Discussion} 

\indent This study was undertaken in effort to constrain the upper-metallicity branch of the \R\ relation with direct-method oxygen abundances.
It was understood from the start that this would be a challenging task.
Galaxies for which the temperature-sensitive \OIII$\lambda$4363 auroral line is strong enough to be readily measured correspond to those of highest excitation and lowest overall metallicity.
The two primary selection criteria used to generate the targets for this project -- strong, high-EW emission lines and high coarse abundances -- are fundamentally at odds with one another:
Our goal was to populate the lower portion of the upper-metallicity branch, above the turnaround region, with robustly obtained abundances and provide observational constraint for the \R\ relation.
We hoped to achieve this goal by (1) utilizing the large light-collecting area of the Keck telescope and the high sensitivity of LRIS, and (2) to pre-select optimal candidates for observation using the best targets from the deep KISS catalogs.\\
\indent Unfortunately, we were unsuccessful in achieving the measurements of high enough metal abundances to populate the upper-metallicity branch of the \R\ relation.
We were, however, successful in deriving reliable direct-method \Te-abundances for the majority of our targets.
Furthermore, we have compared our results to two SEL abundance methods in order to explore the apparent discrepancy in metallicity results, and have explored a non-traditional avenue of utilizing NIR \SIII\ measurements to obtain electron temperatures for our abundance analysis.
Finally, our study has revealed disparity in the homology of an outwardly-appearing homogeneous set of star-forming galaxies.

\subsection{Metallicity Comparison} 

\begin{figure*}
\plotone{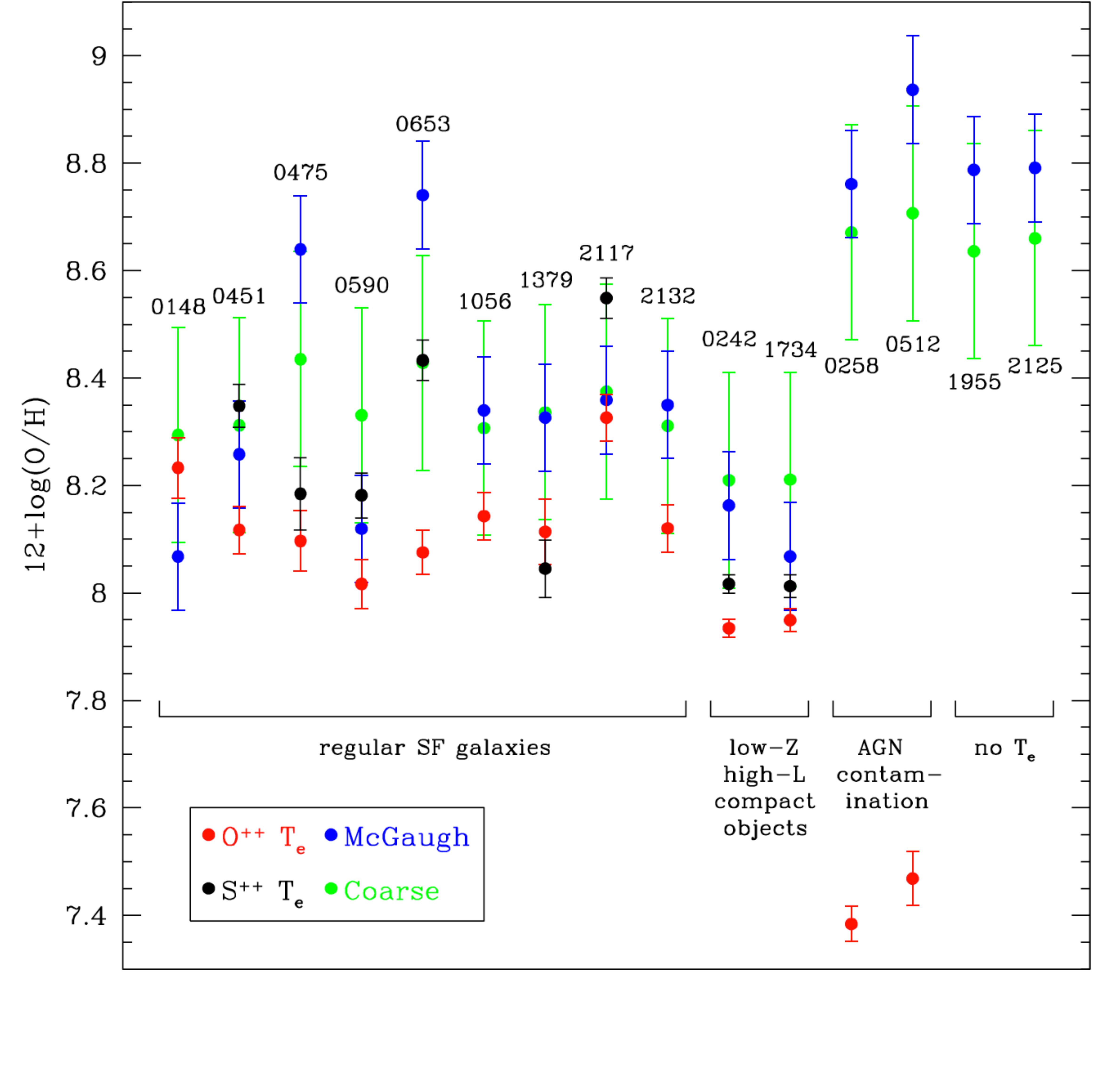}
\vspace{-1.5cm}
\caption{Metallicity comparison for all 15 galaxies from this study utilizing both \OIII\ and \SIII\ for \Te-method abundances (red points and black points, respectively) and the SEL empirical McGaugh model grid (blue points) and coarse abundance method (green points).
For the eight galaxies with calculated S$^{++}$ \Te\ (including six regular star-forming galaxies and two low-metallicity, high-luminosity compact galaxies), the sulfur-derived abundances are higher than oxygen-derived abundances for all but one case.
The \SIII\ lines used to derive the electron temperature represent a different ionization zone in the nebulae than the \OIII\ lines, and when studied in conjunction with traditional \Te-methods provides a reasonable range for the global physical characteristics.
Furthermore, assuming \Te\ calculated from the zone of doubly-ionized sulfur reduces the discrepancy between direct- and SEL methods, though we stress that this technique should be considered a thought experiment in deriving a lower-limit to the overall metallicity.}
\label{fig:abun_plot}
\vspace{0.5cm}
\end{figure*}

\indent Several studies have indicated the existence of discrepancies between oxygen abundances computed via photoionization models and the direct-method.
In such cases, the auroral lines may not produce a reliable \Te\ measurement \citep{bib:Stasinska2005}.
There are strong exponential dependences on the strength of the weak, temperature-sensitive auroral emission lines.
Small-scale electron temperature fluctuations are a likely contributor toward these discrepancies (e.g., \citealp{bib:Peimbert1967, bib:PeimbertCostero1969, bib:Peimbert2003, bib:Peimbert2007}).
In such a case, an unresolved ELG spectrum encompasses a large volume of ionized gas that includes many different individual regions of varying physical properties.
The measured spectrum would therefore comprise a global average that is not representative of any specific region of the nebula.
Detection of \OIII$\lambda$4363 emission would therefore be weighted toward areas of highest excitation and may not necessarily coincide spatially with other regions of weaker \OIII\ emission, where the temperature of the electron gas is lower.
Because these spatially distinct regions of the nebula are measured as part of a single spectrum, their overall physical properties become blended together.\\
\indent Additionally, \citet{bib:Garnett1992} explains that thermal gradients can be responsible for an underestimation of derived abundances from measured electron temperatures.
Observations and photoionization models show that nebulae are not necessarily isothermal and the average \Te\ can vary for different ionic species \citep{bib:Garnett1992}.
In such cases the measured \Te\ is biased high, influencing the estimated metallicities through the emitting regions.
Different ionic species' ionization potentials and radiation hardness gradients with optical depth lead to inhomogeneous nebular structure.
For instance, \citet{bib:Stasinska1980} has demonstrated an increase in electron temperature with increasing radius from the ionizing source for metal-rich nebulae.\\
\indent The study of \citet{bib:AndrewsMartini2013} suggest that O$^{+}$ zone temperatures, often obtained using model-derived relationships involving the measured O$^{++}$ temperatures, tend to be overestimated, leading to underestimates in the O$^{+}$ abundance.
As the overall abundance is assumed as the sum of singly- and doubly-ionized oxygen, the undervalued O$^{+}$ abundance results in a net deficit of the calculated direct-method metallicity.
Recently, discrepancies in nebular abundances derived from electron temperature measurements have been explored with the introduction of the ``$\kappa$-distribution" of electron energies (see \citealp{bib:Nicholls2012}, \citealp{bib:Nicholls2013}, and \citealp{bib:Dopita2013}).
An adoption of model \HII\ regions exhibiting high-energy (suprathermal) excess in electron energies appears to resolve some temperature and metallicity discrepancies as compared to an equilibrium described by a Maxwell-Boltzmann distribution.
The study of \citet{bib:MendozaBautista2014}, however, found that the reconciliation of \Te\ measurements for higher-excitation zones (such as O$^{++}$) using $\kappa$-distributions is not significant.\\
\indent Some studies (notably \citealp{bib:AndrewsMartini2013}) utilize spectra stacked in bins of mass, averaged to achieve the needed signal-to-noise to detect $\lambda$4363.
While such a method does away with certain selection effect biases described above, the disparity between direct- and SEL-method abundance estimations remains.
Spectral stacking, however, may introduce other uncertainties that can skew measurements.
In particular, for a given galaxy mass bin, the inclusion of particularly strong \OIII\ emitters can over-represent $\lambda$4363 emission strength relative to the averaged $\lambda\lambda$4959,5007, biasing temperatures high and thus total oxygen abundances low.\\
\indent With increasing recent evidence that the standard procedure in computing oxygen abundances may be less robust than traditionally assumed, it becomes useful to explore the utility of alternative methods.
A comparison of SEL method metallicities, described in \S5.1.1., as well as a comparison of \Te-method metallicities derived from O$^{++}$ and S$^{++}$, described in \S5.1.2, is shown as Figure \ref{fig:abun_plot}.
The results of our abundance analyses are summarized in Table 4.
We find that oxygen abundances calculated using empirical relations are systematically higher for our targets as compared to direct-methods, and furthermore that metallicities derived from S$^{++}$ \Te\ tend to be higher than as calculated using O$^{++}$ \Te, but are still lower than SEL-method abundances.

\subsubsection{Comparison Between Direct and SEL Abundances} 

\begin{figure*}
\epsscale{0.736}
\plotone{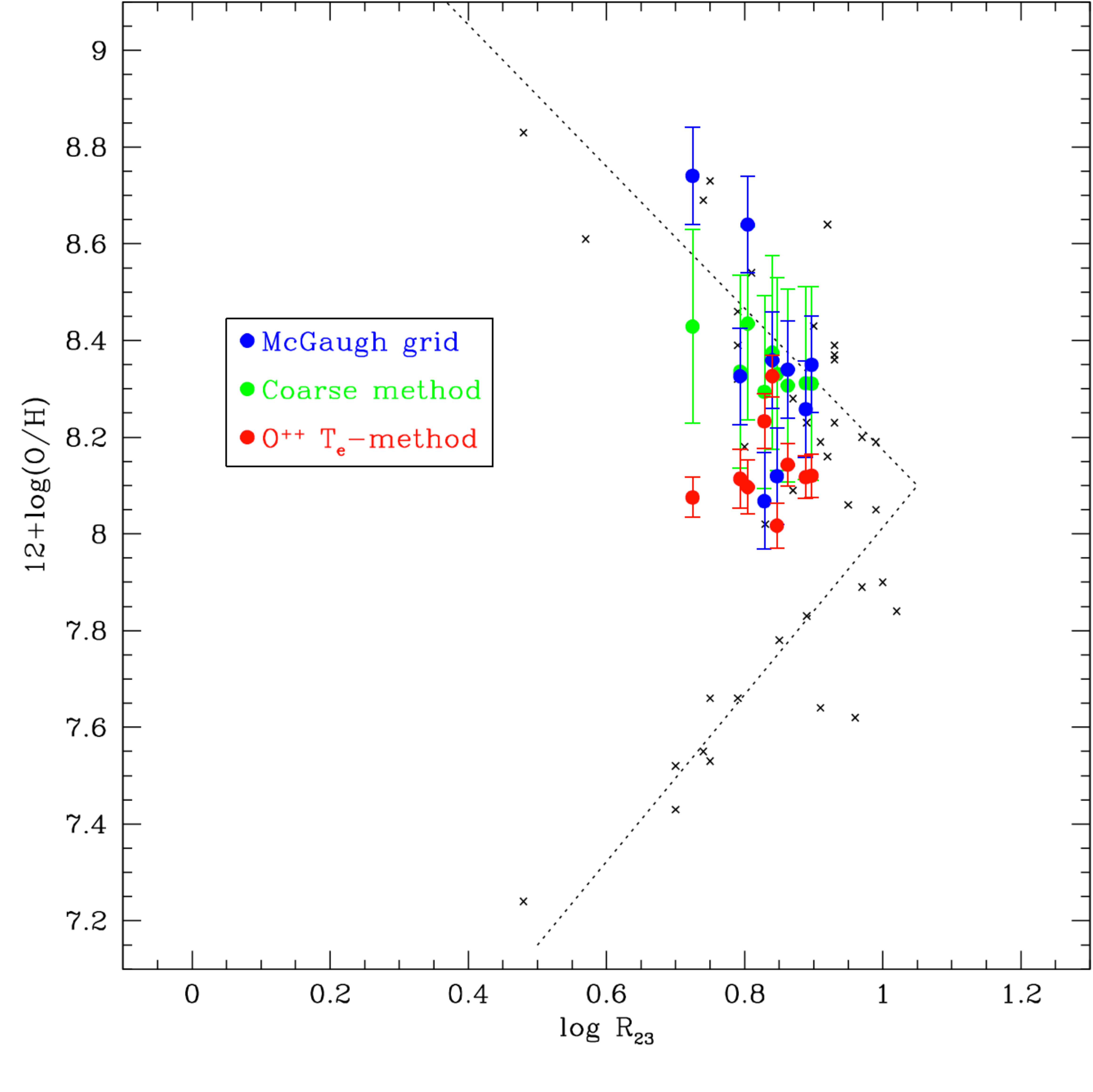}
\vspace{-0.3cm}
\caption{\R\ diagram using oxygen \Te-abundances, McGaugh model grid abundances, and coarse method abundances for our nine regular star-forming galaxies for which \OIII$\lambda$4363 was measured.
The dotted line represents the \R\ calibration from \citet{bib:EdmundsPagel1984}, while the crosses are reference points from \citet{bib:McGaugh1991}.
The SEL-method abundances (McGaugh and coarse) appear systematically high compared to the direct \Te-method.
Note that while some \Te-method values are widely variant from SEL method values, others agree quite well, implying that this offset is not entirely a systematic effect, but can include bias in sample selection.
Errors in line ratio measurements (x-axis) are comparable in size to the dots.
Errors in SEL abundance values (y-axis) are the high-end estimates for either method; $\sigma$ $\approx$ 0.10 for the McGaugh grid and $\sigma$ $\approx$ 0.20 for the coarse method.}
\label{fig:R23_FINAL}
\vspace{0.5cm}
\end{figure*}

\indent Offsets may arise when estimating metallicities for the same object while employing direct- and SEL-methods, where generally \Te-method abundances tend to be biased low when compared to SEL calibrations.
\citet{bib:Kennicutt2003} found that their direct-method metallicities of individual \HII\ regions in M101 were systematically low by 0.2--0.5 dex compared to the most widely used SEL methods.
\citet{bib:Bresolin2009} came to a similar conclusion based on a study of \HII\ regions in NGC 300.
The study of \citet{bib:KewleyEllison2008} found an even greater discrepancy in star-forming galaxies from the Sloan Digital Sky Survey (SDSS; \citealp{bib:Abazajian2005}) of up to 0.7 dex.
\citet{bib:AndrewsMartini2013} argue that the most likely cause of the offset between theoretical- and direct-method metallicities lies in the breakdown in one or more of the assumptions in the physics of \HII\ regions in the stellar synthesis or photoionization models.
Stellar population synthesis models typically assume that the ionizing source is a zero age main sequence starburst, which affects line fluxes as a function of age \citep{bib:Berg2011}.
Additionally, photoionization models typically invoke simplifying assumptions or treatments that reduce their applicability to real-life nebulae.
These include the patently false assumption of spherical or plane-parallel \HII\ region geometry, poorly constrained treatment of metal depletion onto dust grains (e.g., \citealp{bib:ZuritaBresolin2012} and references therein), and the assumption of smooth (rather than clumpy) gas and dust density distribution \citep{bib:KewleyEllison2008}.\\
\indent In order to quantify the amount of offset between direct- and SEL-methods and the effect this has on the calibration of the \R\ relation, a comparison of three abundance estimates for the nine regular star-forming galaxies in this study with measured \OIII$\lambda$4363 are plotted with their respective \R\ ratios as Figure \ref{fig:R23_FINAL}.
Using the McGaugh model grid, the estimated oxygen abundance values appear to be systematically high by an average of 0.22 $\pm$ 0.25 dex as compared to those calculated using the \Te-method.
Metallicities estimated using the coarse abundance method also tend to be high relative to the \Te\ abundances with an average offset of 0.21 $\pm$ 0.11 dex.
There is zero average offset between the SEL McGaugh and coarse abundance methods for the nine regular star-forming galaxies, but with a standard deviation of 0.17 dex.
With a few exceptions, the SEL method abundances are greater than \abun\ = 8.2 and fall on the upper-branch of the \R\ metallicity diagram, while direct-method abundances calculated \emph{from the same data} remain in the turnaround region.\\
\indent The two primary sample selection criteria for this study specify both high-EW \OIII\ lines and potentially high oxygen abundances as estimated using the SEL coarse abundance method.
Galaxies selected from KISS that satisfy the necessity for strong \OIII\ emission will be high-excitation and have higher electron gas temperatures, indicative of inefficient cooling and thus must be relatively metal-poor.
The satisfaction of the selection criterion mandating high potential metallicities for these galaxies, as estimated by the coarse method, therefore introduces a selection bias.
We believe that the viable targets initially picked for this study are specifically those whose coarse-method abundance estimations occupy the high end of the scatter of the metallicity distribution.
In the majority of cases, as seen in Figure \ref{fig:R23_FINAL}, the different metallicity estimates agree within their respective errors, with SEL method values occupying the high side of the \Te-method scatter.
Metallicity estimates for a few of the regular star-forming galaxies, namely KISSR 2117, agree quite well between the direct- and SEL methods.
This provides evidence that any offset present is one that is not entirely systematic.

\subsubsection{Comparison Between O$^{{\text +}{\text +}}$ and S$^{{\text +}{\text +}}$ \Te\ Abundances} 

\indent We have derived the electron temperature for several of the ELGs from this study using \SIII\ lines (\S4.1.2).
For all but one of these objects, the sulfur \Te\ was lower than that of oxygen, however the two temperatures agree within the errors.
All eight of these galaxies are considered star-forming galaxies (e.g., none are thought to possess spectral contamination by AGN), though two fall into the classification of high-luminosity, low-metallicity compact objects (see \S5.2.3).\\
\indent It is possible to compare the oxygen abundance \abun\ as calculated using electron temperatures derived from both oxygen and sulfur.
When the temperature-sensitive auroral \SIII$\lambda$6312 line is measured in ELG spectra, the S$^{++}$ \Te\ is usually used only to compute the sulfur abundance.
The S$^{++}$ electron temperature is representative of a slightly lower-energy ionization zone within the nebulae as compared to O$^{++}$, and as such the \Te\ of the S$^{++}$ zones are not necessarily appropriate substitutes for those of the O$^{++}$ zones.
In our study, however, we have opted to explore a \emph{non-standard method} of deriving oxygen abundances by employing the electron temperature estimated from the S$^{++}$ zone.
Utilizing sulfur as the nebular thermometer for abundance calculations may reduce certain biases in oxygen measurement (see \S4.1.2.).\\
\indent The adoption of the S$^{++}$ \Te\ for the temperature of the high-ionization O$^{++}$ zone gives a range in physically viable values for the calculation of metallicities.
Implementing abundance estimates utilizing sulfur-derived electron temperatures should not be taken as the solution to oxygen measurement biases, but rather as a complementary method.
Abundances using S$^{++}$ \Te\ are correspondingly higher compared to the traditional method for the same galaxies.
These values fall between the O$^{++}$ \Te-method and SEL abundance estimates, potentially offering some reconciliation of the disparity discussed in \S4.3.1.
We note that we chose not to include the S$^{++}$ \Te-substituted oxygen abundance for KISSR 2125 in our analysis because the resulting error bars on the estimated value, \abun\ = 8.06 $\pm$ 0.41, are unreasonably large (more than double the error of the coarse method abundance).\\
\indent Regular star-forming galaxies for which we obtained NIR spectral coverage have direct-method metallicities estimated to be higher by an average of 0.16 $\pm$ 0.14 dex.
Should the S$^{++}$ ionization zone temperature be more representative of the nebula as a whole, then the metallicity derived using \OIII\ emission lines will underestimate the global oxygen abundance.
This small, systematic offset between O$^{++}$ and S$^{++}$ \Te-method oxygen abundances reduces the average overall discrepancy between direct- and SEL-method metallicities seen in Figures \ref{fig:abun_plot} and \ref{fig:R23_FINAL}.
Adopting the oxygen abundances of galaxies estimated using sulfur temperatures reduces the average offset between direct-method and SEL-method metallicities from 0.26 $\pm$ 0.22 to 0.11 $\pm$ 0.22 dex (for the McGaugh model grid) and from 0.25 $\pm$ 0.10 to 0.11 $\pm$ 0.16 dex (for the coarse method) for our targets that possess NIR spectral coverage.
Because their origin is from a lower-energy transition, S$^{++}$ zone-derived electron temperatures could therefore be less subject to the effects of temperature fluctuations than O$^{++}$ on the derivation of direct-method metallicities.\\
\indent \citet{bib:Esteban2014} show that within their study, giant \HII\ regions in dwarf star-forming galaxies exhibit the highest temperature fluctuation parameters.
As the measured discrepancy between direct- and SEL-method abundances is thought to be at least in part a consequence of such fluctuations, the adoption of sulfur-derived electron temperatures in direct-method analyses could produce intermediate metallicity estimates useful in reconciling the standard techniques.
While some of the discrepancy between direct- and SEL-method abundance calculations appears to be accounted for by utilizing S$^{++}$ rather than O$^{++}$ electron temperatures, we do not prescribe that such a surrogate is the best solution for deriving oxygen abundances from ELG spectra.
Rather, this method can be used as a constraint to direct-method metallicities by providing a reasonable range of electron temperature values for use in \Te-method abundance determinations.\\
\indent For a galaxy or \HII\ region in which an O$^{++}$ \Te\ cannot be derived, measurements of the \SIII\ emission lines may enable an alternative direct-method metallicity.
For higher-metallicity nebulae, \SIII$\lambda$6312 will be the preferably-detected auroral line as the strength of \OIII$\lambda$4363 diminishes \citep{bib:Croxall2015}.
The S$^{++}$ temperature could then be used to estimate an O$^{++}$ temperature, from which an oxygen abundance could be found.
Further observations and analysis of targets with NIR spectral coverage and S$^{++}$ \Te\ determinations will provide additional data points from which to refine this offset.

\subsection{Inhomogeneity of Sample Galaxies} 

\begin{figure*}
\plotone{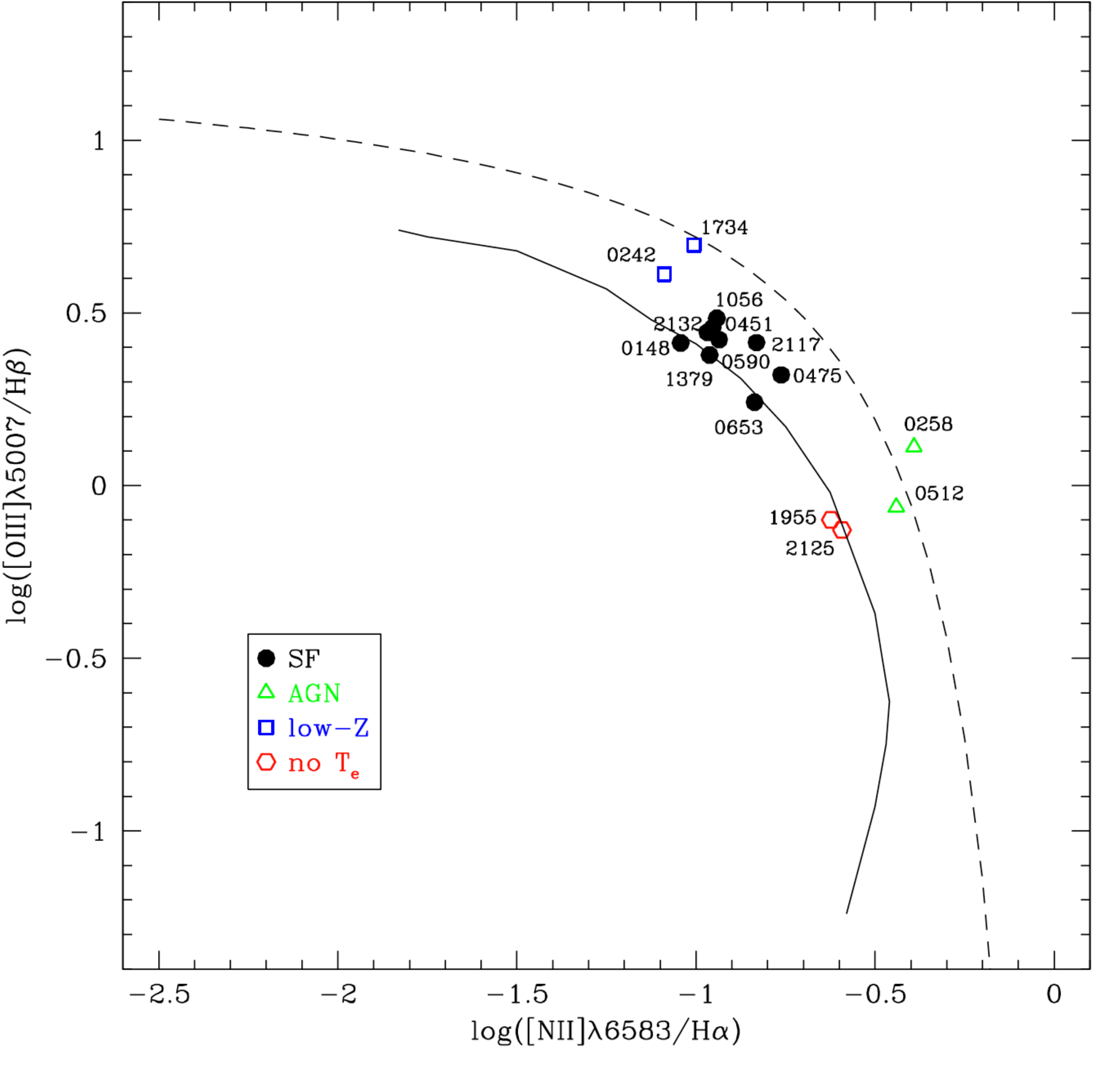}
\vspace{-0.5cm}
\caption{Diagnostic diagram of galaxies we observed with Keck segregated by type, including regular star-forming galaxies for which \Te-abundances have been derived in this study (black filled circles), those with no believable detections of the temperature-sensitive \OIII$\lambda$4363 \AA\ line and thus lower precision abundance measurements (red open hexagons), those with suspected contamination by a non-thermal photoionizing source, e.g., AGN (green open triangles), and possible nearby examples of low-metallicity, high-luminosity galaxies (blue open squares).
The line ratios plotted here are those found using Keck LRIS from this study.}
\label{fig:Keck_DD_labelsP}
\vspace{0.5cm}
\end{figure*}

\indent When we selected our objects for inclusion in this study, we tacitly assumed that all of the KISS galaxies in our lists were typical star-forming galaxies.
After obtaining and analyzing our spectra, however, it became clear that not all of our sources could be categorized as simple or ``regular" star-forming systems.
That is, the targets of this study appear to compose an inhomogeneous set of objects.
We have segregated our galaxies into three distinct categories:  Regular star-forming galaxies (further distinguished into those with and without \OIII$\lambda$4363 detection), those with suspected AGN spectral contamination, and low-metallicity, high-luminosity compact objects.
We discuss each of these subgroups below.
Figure \ref{fig:Keck_DD_labelsP} presents a line-ratio diagnostic diagram plotting the newly measured ratios and using different symbols to indicate the four classes of objects in our sample.

\subsubsection{Regular Star-Forming Galaxies} 

\indent Eleven of the 15 targets were determined to be ``regular" star-forming galaxies, for which \Te-method abundances could be calculated for nine (represented by filled circles in Figure \ref{fig:Keck_DD_labelsP}).
Of the latter group, the calculated metallicities all inhabit a range of 8.0 $\lesssim$ \abun\ $\lesssim$ 8.3, and therefore were not successful in extending above the \R\ turnaround regime and populating the upper-metallicity branch, as seen in Figure \ref{fig:R23_FINAL}.
These \Te-method abundances are on average 0.21 $\pm$ 0.12 dex lower than the initial coarse abundance estimates from the original KISS follow-up spectroscopy.
For some galaxies, such as KISSR 2117, the direct-method metallicities agree quite well with the initial coarse method estimates, with \abun\ = 8.33 and 8.36, respectively.
Others, such as KISSR 0653, are widely discrepant, with \abun\ = 8.08 and 8.46, respectively.
Despite the fact that we were unsuccessful in determining direct abundances above the turnaround region, we have provided robust measurements of the oxygen abundance for several KISS galaxies.
These will be added to the overall sample of KISS galaxies possessing direct abundances and will be useful, for example, in future refinements to the calibration of the SEL coarse abundance method.\\
\indent The two remaining star-forming galaxies, KISSR 1955 and KISSR 2125, lacked any discernible detection of the \OIII$\lambda$4363 temperature-sensitive line (represented by open hexagons in Figure \ref{fig:Keck_DD_labelsP}).
Therefore we were unable to calculate a \Te-method oxygen abundance.
While appearing to be regular star-forming galaxies in all other respects, the initial abundance estimate for these two targets were amongst the highest of the sample (\abun\ = 8.60 and 8.65, respectively), and so it should not be surprising that they represent the technical limit of our ability to recover metallicities via the direct-method.
Lacking a measurement of a direct-method oxygen abundance, these two galaxies were eliminated from further analysis in this sample.

\subsubsection{AGN Contamination} 

\indent Two of the 15 targets, KISSR 0258 and KISSR 0512, appear to possess emission-line ratios indicative of spectral contamination by a non-thermal ionizing source (represented by open triangles in Figure \ref{fig:Keck_DD_labelsP}).
Both of these galaxies appear near or past the \citet{bib:Kauffmann2003} boundary for AGN activity, despite initially appearing to be regular star-forming galaxies (Figure \ref{fig:KoK_DD}) by the original KISS spectral data.
A direct comparison with Figure \ref{fig:KoK_DD} furthermore indicates that while the initial position of KISSR 0512 is nearby its location on Figure \ref{fig:Keck_DD_labelsP}, there appears to be a major discrepancy in the position of KISSR 0258.
While initially believed to be a regular star-forming galaxy, this target has clearly shifted beyond the demarcation line segregating AGN.
A comparison with the initial KISS follow-up spectroscopy implies that the original pointings of this object placed the slit over a different region of the galaxy than was observed by Keck, e.g., an outlying knot of star formation compared to nuclear emission.
Based on the location of this target on the diagnostic diagram of Figure \ref{fig:Keck_DD_labelsP}, it should not have been included as a potential candidate for observation by this study.
The location of KISSR 0512 on the diagnostic diagrams depicting original KISS data and our values from Keck (Figures \ref{fig:KoK_DD} and \ref{fig:Keck_DD_labelsP}, respectively) resides near the boundary separating star-forming galaxies and AGN and therefore was already suspect.\\
\indent As a consequence of their placement on the diagnostic diagram, we feel that these two galaxies are likely contaminated by AGN activity, and hence cannot be counted amongst the regular star-forming galaxies.
In particular, their detected \OIII$\lambda$4363 lines appear to be far too strong for a purely star-forming emission-line source.
These AGN-contaminated targets have major offsets due to the non-thermal nature of the photoionization sources powering the nebulae.
Because of the high-energy photons emitted by matter accreting onto the galaxy's central supermassive black hole, there is a comparatively harder ionizing radiation field that disproportionally inflates the emission of \OIII$\lambda$4363 in relation to stronger lines such as \OIII$\lambda\lambda$4959,5007.
As a result, the temperature-diagnostic emission-line ratio produces a measured electron temperature that is skewed high.\\
\indent Consequently, these two galaxies present the most disparate values of metallicity between the direct- and SEL methods.
\Te-method oxygen abundances for KISSR 0258 and KISSR 0512 were calculated to have values of \abun\ = 7.38 and 7.47, respectively, as seen in Table 4.
These values are extremely low and are therefore probably spurious.
Especially considering that these two galaxies are the most luminous in our sample ($M_{B}$ = -20.15 and -20.46, respectively), their low measured metallicities are very surprising when placed on the luminosity-metallicity ($L$--$Z$) relation.
The disparity of the \Te\ abundances compared to their coarse method estimates (\abun\ = 8.41 and 8.86, respectively) is quite drastic.
For such cases, the physical assumptions involved with the star-formation models that are integral to the development and calibration of SEL method relations are no longer directly applicable.

\subsubsection{Low-Metallicity, High-Luminosity Compact ELGs} 

\begin{figure*}
\plotone{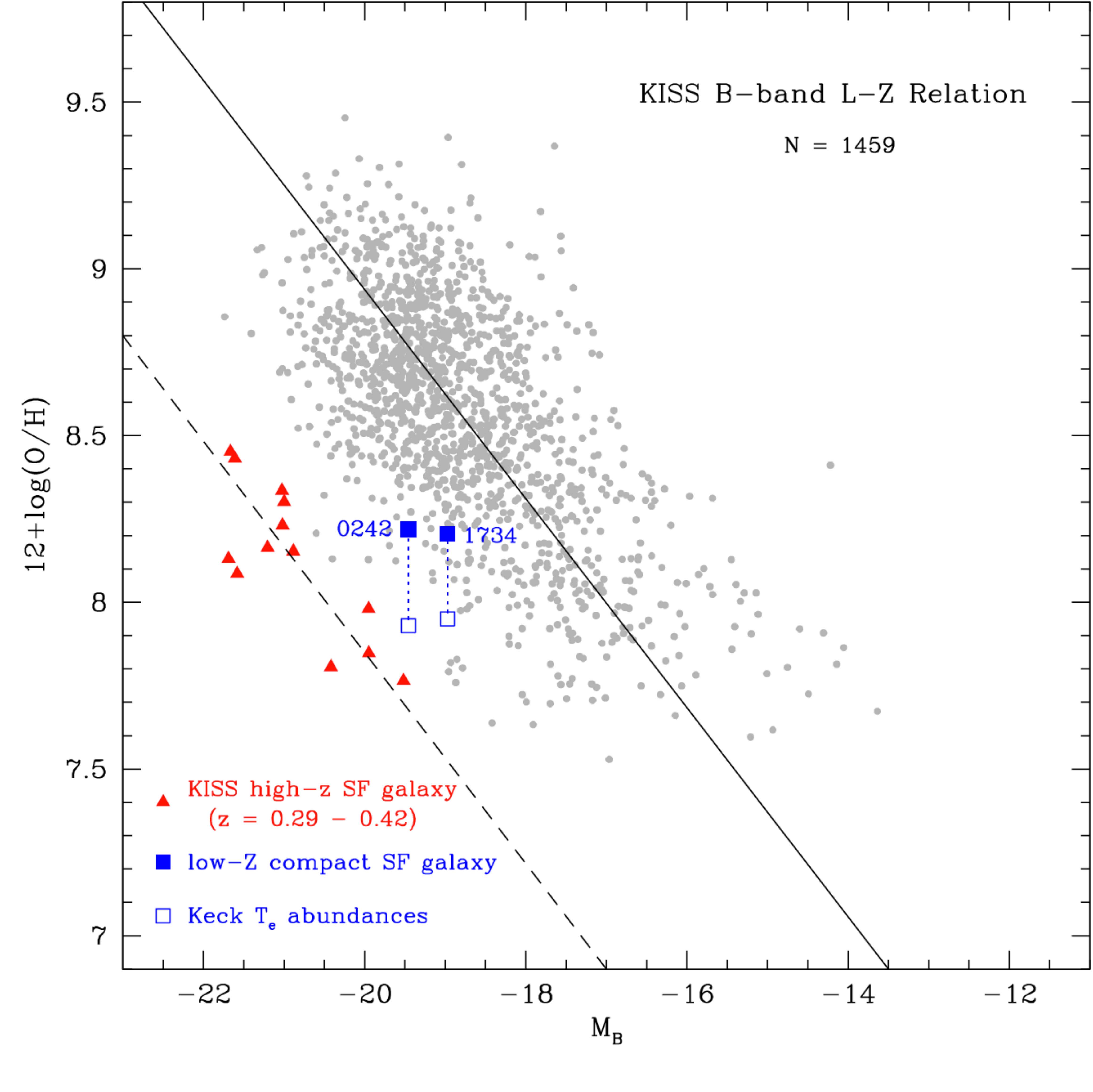}
\vspace{-0.5cm}
\caption{Luminosity-metallicity ($L$-$Z$) relation diagram for 1400+ low-$z$ KISS regular star-forming galaxies (small dots) adopted from \citet{bib:Salzer2009} with 13 \OIII-detected star-forming galaxies (red triangles).
Solid blue squares are the two possible local analogues of low-metallicity, high-luminosity compact galaxies discovered by this study.
Metallicities are estimated using the coarse abundance method.
Open blue squares represent \Te-method abundances calculated in this study, connected to initial metallicity estimates with a dotted line.
These abundances are more than 2$\sigma$ outliers to the norm.
The solid line is a linear fit to the low-$z$ galaxies, while the lower dashed line has the same slope but fits the higher-$z$ galaxies with an offset of -1.1 dex.}
\label{fig:KISS_L-Z}
\vspace{0.5cm}
\end{figure*}

\indent The remaining two targets, KISSR 0242 and KISSR 1734, both exhibit very strong \OIII$\lambda$4363 emission.
Both were found to be highly underabundant (\abun\ = 7.93 and 7.95, respectively), but also exhibit very compact morphologies and high luminosities ($M_{B}$ = -19.46 and -18.97, respectively).
This combination contrasts with the well-established $L$--$Z$ relation, whereby the most luminous galaxies possess the greatest degree of enrichment.
We note that the direct-method metallicities for these objects relied on \Te\ measurements utilizing very high signal-to-noise spectra, owing to the high-excitation nature of these galaxies and the exceptionally strong emission of \OIII$\lambda$4363 (see Figures \ref{fig:block1} and \ref{fig:block3}).
As a result, the oxygen abundances calculated for these galaxies are very robust.\\
\indent Galaxies of this nature have been previously detected in the KISS sample \citep{bib:Salzer2009}, however those objects were all detected as strong \OIII-lined sources redshifted such that they were detected in the KISS \HA\ filter.
The redshift range of these rare \OIII-detected KISS galaxies is $z$ = 0.29 -- 0.42, corresponding to a lookback time of roughly 3--4 Gyr (assuming a cosmology of $H_{\emph{0}}$ = 70 km s$^{-1}$, $\Omega_{M}$ = 0.27, and $\Omega_{\Lambda}$ = 0.73) compared to a redshift range of $z$ = 0.04 -- 0.07 for our two targets from the current study.
To match what is typically observed in the local Universe today, these higher-redshift \OIII-detected KISS galaxies would have necessarily been required to experience an immense degree of metal enrichment (a factor of $\sim$10) or to have undergone a massive luminosity decrease (a factor of $\sim$28).
Either of these scenarios show that such luminous, metal-poor galaxies represent extreme phases of evolution.
As shown in \citet{bib:Salzer2009}, these galaxies are very rare, representing less than 0.02\%\ of all galaxies in this luminosity range ($M_{B}$ from -19.5 to -22.0) at these lookback times.\\
\indent Figure \ref{fig:KISS_L-Z} illustrates the location of these higher-redshift objects in relation to the set of regular star-forming KISS galaxies on an $L$--$Z$ diagram (metallicities estimated using the coarse abundance method), and includes these two possible local analogues.
Solid squares represent initial abundance estimates, demonstrating that these objects are metal-poor and/or over-luminous relative to typical KISS galaxies.
The degree by which they are discrepant from the rest of the KISS sample, however, is fairly moderate:
Their placement on the outside edge of the main cluster of data points puts them roughly half way between the $L$--$Z$ relation fit and the offset fit to higher-redshift KISS star-forming galaxies from \citet{bib:Salzer2009}.
Utilizing coarse abundance estimates, then, these two objects were not expected to be extreme outliers.
The open squares connected via dotted lines represent \Te-abundances measured from this study, indicating that, modulo uncertainties in the coarse abundance method, KISSR 1734 is a 2$\sigma$ outlier and KISSR 0242 is nearly a 3$\sigma$ outlier relative to the mean $L$--$Z$ relation for the full KISS sample.\\
\indent Other recent studies, including \citet{bib:Hoyos2005}, \citet{bib:Kakazu2007}, and the ``Green Pea" galaxies (GPs) introduced by \citet{bib:Cardamone2009}, have also reported discovery of luminous, underabundant galaxies in various redshift ranges.
A study by \citet{bib:Amorin2010} shows that GPs are found to follow a mass-metallicity ($M$--$Z$) relation parallel to regular star-forming galaxies, but one that is offset by $\lesssim$ 0.3 dex to lower metallicities.
These objects are therefore similar but less extreme in their divergence from typical galaxies than those from \citet{bib:Salzer2009}.
Potentially, GPs and their equivalents are a subset of star-forming luminous compact galaxies (LCGs).
This idea demonstrates that such objects exist at any redshift regime, but present differing observable properties inducing classification as a variety of objects \citep{bib:Izotov2011}.
\citet{bib:Cardamone2009} provides evidence that GPs/LCGs are nearby analogues of starbursting UV-luminous galaxies (UVLGs) such as Lyman-break galaxies (LBGs) and Ly$\alpha$ emitting galaxies (LAEs).
This inference is verified by UV spectra of some GPs showing Ly$\alpha$ emission \citep{bib:JaskotOey2014}.
Star-formation taking place within UVLGs in a redshift range 1.9 $\textless$ $z$ $\textless$ 3.4 represents a relatively large portion of the current total stellar mass \citep{bib:ReddySteidel2009}.
GPs/LCGs therefore represent relatively nearby laboratories with which to study the mechanisms at work consistent with the bulk of star-formation in the Universe.\\
 \indent Another related class of galaxies are the luminous blue compact galaxies (LBCGs) \citep{bib:Garland2004, bib:Werk2004}.
These objects possess luminosities equivalent to or greater than the Milky Way, and yet are substantially more compact.
Though common at intermediate and high redshifts, LBCGs are locally rare.
With activity dominated by intense star formation, these blue galaxies contributed significantly to the overall star-formation rate density of the Universe.
The study of \citet{bib:Werk2004} examined 16 LBCG candidates selected from KISS, including KISSR 0242.
They found that LBCGs have higher \HA\ luminosities than average, indicative of strong star-formation activity, and lower metal abundances than expected based on the $L$--$Z$ relation for KISS galaxies.\\
\indent The import of finding low-metallicity, high-luminosity star-forming galaxies in the very local Universe is one that speaks to the ongoing understanding of how these systems evolve.
Objects that are divergent low from the $L$--$Z$ relation, like those at $z$ = 0.29 -- 0.42 discussed in \citet{bib:Salzer2009}, are implied to have experienced significant chemical enrichment over the subsequent $\sim$4 Gyr in order to evolve into more typical star-forming systems at $z$ = 0.
These galaxies are thought to be amongst the final massive systems in the Universe that are in an early phase of gravitational collapse and initiation of star formation.
But finding local representatives of these objects ($z$ = 0.038 and 0.066 for KISSR 0242 and KISSR 1734, respectively) indicates that such an intense initial evolutionary stage may have been delayed further until even very recently.\\
\indent That two galaxies of this type were found from a set of only 15 is not an indication that these are common.
On the contrary, of initial the list of potential KISS targets, KISSR 0242 and KISSR 1734 possessed the highest \OIII$\lambda$5007 equivalent widths.
Furthermore, their estimated oxygen abundances and absolute magnitudes were also amongst the lowest of all listed galaxies.
Because the selection criterion of this study mandated strong \OIII\ emission, it should be no surprise that low-metallicity, high-luminosity targets would be preferentially selected from the overall sample for observation.
The fact that a handful of additional local galaxies from our target list possessing similar qualities have not yet been observed leaves the exciting prospect that more systems in an equivalent evolutionary stage are available for further study.
Future work promises to shed new light on this type of system.

\section{Summary and Conclusions} 

\indent We have presented spectral analysis of 15 ELGs observed with Keck LRIS and selected from KISS to be both strong \OIII\ emitters and metal-rich in effort to find \Te-method abundances to populate and constrain the upper-metallicity branch of \R.
We were successful in recovering \Te-method abundances for 13 of the 15 galaxies observed.
An examination of the spectra and their location on a diagnostic diagram reveals that the spectral activity of our targets are inhomogeneous, including two objects with probable AGN contamination and two compact luminous galaxies with low abundance.
Oxygen abundances calculated for nine regular star-forming galaxies of this study were found to be constrained to the turnaround region of the \R\ diagram, corresponding to 8.0 $\lesssim$ \abun\ $\lesssim$ 8.3.\\
\indent Of the three lists of initial selection criteria, galaxies from the third list (modest \OIII\ EW and highest estimated abundance) were bereft of measurable $\lambda$4363, and therefore of a \Te-method abundance measurement.
The second list (moderate \OIII\ EW and moderate estimated abundance) produced galaxies with direct-method metallicities firmly in the \R\ turnaround region, a useful determination as a robust value of oxygen abundance but falling short of the goals for this study.
The first list (highest \OIII\ EW and moderate estimated abundance), however, selected galaxies that we either identified as low-metallicity, high-luminosity compact star-forming galaxies (and are therefore interesting as a separate case) or as regular star-forming galaxies, including some with fairly high measured oxygen abundances.\\
\indent The highest metallicity we found overall was for a galaxy from the first list, KISSR 2117, for which we calculated an oxygen abundance of \abun\ = 8.33 $\pm$ 0.04.
While not enriched enough to populate the upper-metallicity branch of \R, the abundance calculated for this galaxy is approaching the necessary limit to do so.
To fulfill our original project goal, we suggest that searches for targets of future observations should focus on the strength of \OIII$\lambda$5007, in order to recover $\lambda$4363 for determination of \Te.\\
\indent We have also reaffirmed the results of previous studies that in comparison to the \Te-method, the McGaugh model grid and coarse abundance empirical strong-line methods result in oxygen abundance values that are systematically higher (by 0.22 $\pm$ 0.25 dex and 0.21 $\pm$ 0.11 dex, respectively).
While typically thought to be a result of certain well-known observational biases such as temperature fluctuations and gradients, in this case such disparity is likely enhanced by a bias in the selection criteria employed for our study.
Preliminary metallicity estimates based on the coarse abundance method, for objects with strong \OIII\ emission, will be preferentially located in the high-end scatter of the distribution, and as such the discrepancies between our direct- and SEL-method abundances may be significantly enhanced.\\
\indent A comparison of electron temperature calculations utilizing emission lines of O$^{++}$ and S$^{++}$, representing overlapping ionization zones in the nebulae, yield systematic differences in direct-method oxygen abundance values, a result consistent with previous studies.
For the eight cases in which all required lines were measured in the galaxy spectra, all but one showed overall lower temperatures when measured with S$^{++}$ lines, corresponding to an increase in measured metallicity of an average of $\sim$0.14 dex.
While the adoption of a \Te\ derived from sulfur for abundance calculations cannot be taken as the solution to oxygen measurement biases, a comparison with standard procedures (e.g., \Te\ calculated from oxygen) provides a complementary value.
The resulting lower measured electron temperatures correspond to higher estimated abundances, potentially offering some reconciliation of the disparities between direct- and SEL-method metallicities.
Whether this is because sulfur-derived \Te\ is a better representation of the true nebular temperature or not is unknown.
Furthermore, with a standardized temperature offset, it is possible to estimate an electron temperature for a zone of doubly-ionized oxygen from the measurement of emission lines of doubly-ionized sulfur.
This is potentially very useful for observations of targets where reddening is a major impediment or blue sensitivity is unavailable.\\
\indent Based on analysis of the spectral activity type derived from the diagnostic diagram (Figure \ref{fig:Keck_DD_labelsP}), we have concluded that our targets comprise an inhomogeneous set of galaxies.
The majority of objects (11 of 15) are regular star-forming galaxies, albeit for two (KISSR 1955 and KISSR 2125) we were unable to measure the temperature-sensitive \OIII$\lambda$4363 auroral line and thus recover direct-method abundances.
Two of our galaxies (KISSR 0258 and KISSR 0512) demonstrate elevated line-ratios indicative of non-thermal photoionizing sources, and are thus likely to be contaminated by AGN activity.
Finally, two of our galaxies (KISSR 0242 and KISSR 1734) appear to be under-abundant for their luminosities, falling below the $L$--$Z$ relation defined by the KISS sample.
They appear to represent very local analogues of the low-metallicity, high-luminosity compact objects of \citet{bib:Salzer2009} and the ``Green Peas" found in the SDSS \citep{bib:Cardamone2009}, with implication that they are exhibiting an extreme phase of galaxy evolution not yet heavily studied in the nearby Universe.
That several additional local galaxies from our initial target list were left unobserved leaves the exciting prospect that more such systems are available for study.
\\
\\
\indent We gratefully acknowledge financial support for the KISS project from an NSF Presidential Faculty Fellow to J. J. S. (NSF AST 95-53020), as well as additional support for the ongoing follow-up spectroscopy campaign (NSF AST 00-71114).
The quality of this manuscript was greatly improved by the many useful suggestions made by an anonymous referee.
We thank the many KISS team members who have participated in the spectroscopic follow-up observations.
Finally, we wish to thank the support staff of the W. M. Keck Observatory for their excellent assistance in obtaining the spectroscopic observations that made this work possible.

\bibliographystyle{apj}
\bibliography{Bibliography}

\end{document}